\documentclass[12pt]{amsart}
\usepackage{graphicx,graphics,subfigure}
\usepackage{amssymb}
\usepackage{url}

\usepackage{multicol}

\newtheorem{finding}{Finding}

\pdfoutput=1

\begin{document}

\author{Andrew Adamatzky}
\address[Andrew Adamatzky]{University of the West of England, Bristol, UK} 
\author{Selim G. Akl}
\address[Selim G. Akl]{School of Computing,
Queen's University,
Kingston, Ontario, Canada}

\title[Slime mould imitates Canadian highways]{Trans-Canada Slimeways: Slime mould imitates the Canadian transport network}

\begin{abstract}
Slime mould \emph{Physarum polycephalum} builds up sophisticated networks to transport nutrients between 
distant part of its extended body.  The slime mould's protoplasmic network is optimised for maximum coverage of nutrients yet minimum energy spent on transportation of the intra-cellular material. In laboratory experiments with 
\emph{P. polycephalum} we represent Canadian major urban areas with rolled oats and inoculated slime mould 
in the Toronto area. The plasmodium spans the urban areas with its network of protoplasmic tubes. We uncover similarities and differences between the protoplasmic network and the Canadian national highway network, analyse the networks  in terms of proximity graphs and evaluate slime mould's network response to contamination. 

\vspace{0.5cm}

\end{abstract}

\maketitle

\section{Introduction}

The increase of long-distance travel and subsequent reconfiguration of vehicular  and social networks~\cite{larsen}
requires novel and unconventional approaches towards analysis of dynamical processes in complex 
transport networks~\cite{barrat_2008},  routing and localisation of vehicular networks~\cite{Olariu},
optimisation of interactions between different parts of a transport network during scheduling road expansion and 
maintenance~\cite{taplin}, and shaping of transport network structure~\cite{beuthe}. In the present paper we attempt to build viable analogies between biological and human-made transport networks and project behavioural traits
of biological networks onto existing vehicular transport networks.

While choosing a biological object we want it to be experimental laboratory friendly, easy to cultivate and handle, 
and convenient to analyse its behaviour. Ants would indeed be the first candidate, and a great deal of impressive results has been published on ant-colony inspired computing~\cite{dorigo,solnon}, however ant colonies require 
substantial laboratory resources, experience and time in handling them. Actually very few, if any, papers were published on experimental laboratory implementation of ant-based optimisation,  the prevalent majority of publications being theoretical. There is however an object which is extremely easy to cultivate and handle, and which exhibits remarkably good foraging behaviour and development of transport networks. This is the plasmodium of \emph{Physarum polycephalum}.

Plasmodium is a vegetative stage of acellular slime mould \emph{P. polycephalum}, 
a syncytium, that is, a single cell with many nuclei, which feeds on microscopic particles~\cite{stephenson_2000}.
The plasmodium is a unique user-friendly biological substrate from which experimental prototypes of 
massive-parallel amorphous biological computers are designed~\cite{PhysarumMachines}.  During its foraging behaviour the plasmodium spans scattered sources of nutrients with a network of  protoplasmic tubes. The protoplasmic network is optimised to cover all sources of food and to provide a robust and speedy transportation of nutrients and metabolites in the plasmodium body. The plasmodium's foraging behaviour can be interpreted as computation. 
Data are represented by spatial configurations of attractants and repellents, and  results of computation by structures of a protoplasmic network formed by the plasmodium on the data 
sets~\cite{nakagaki_2000, nakagaki_2001a, PhysarumMachines}. The problems solved by plasmodium of \emph{P. polycephalum} include shortest path~\cite{nakagaki_2000, nakagaki_2001a}, 
implementatiton of storage modification machines~\cite{adamatzky_ppl_2007},
Voronoi diagram~\cite{shirakawa},  Delaunay triangulation~\cite{PhysarumMachines}, 
logical computing~\cite{tsuda_2004}, and 
process algebra~\cite{schumann_adamatzky_2009}; see overview in~\cite{PhysarumMachines}.

Previously~\cite{adamatzky_UC07} we have evaluated a road-modeling potential of \emph{P. polycephalum},
however no conclusive results were presented back in 2007.  A step forward, namely, biological-approximation,
or evaluation, of human-made road networks was done in our previous papers on approximation
of motorways/highways in the United Kingdom~\cite{adamatzky_jones_2009},
Mexico~\cite{adamatzky_Mexico} and the Netherlands~\cite{adamatzky_Netherlands} by plasmodium of  \emph{P. polycephalum}. For all three countries we found that, in principle, the network of protoplasmic tubes developed by plasmodium matches, at least partly, network of human-made transport arteries. The shape of a county and the exact spatial distribution of urban areas (represented by source of nutrients) may play a key role in determining the exact structure of the plasmodium network.  Also  we suspect that a  degree of matching between Physarum networks and motorway networks is determined by original government designs of motorways in any particular country. This is why it is so important to collect data on development of plasmodium networks in all major countries, and then undertake a comparative analysis.

What  are unique properties  of the Canadian transport system? The Canadian Highway System gives us a good example of a logically designed transportation system whose  key goal is to connect all the provinces together by highways. The highway network was  built as a federal-provincial territorial cooperative effort with great effort taken in coordinating work on different parts. Another attractive property of the highway system is that it was designed to provide an access to remote areas where no spots of high population density exist.  

The paper is structured as follows. We present experimental techniques used in Sect.~\ref{methods}. Properties of protoplasmic networks built by \emph{P. polycephalum} are discussed in Sect.~\ref{physarumgraphs}. We compare 
slime mould networks with the Canadian highway network in Sect.~\ref{PhysarumVsHighways} and slime mould and human-made networks with proximity graphs in Sect.~\ref{proximitygraphs}. In Sect.~\ref{contamination} we show 
how the slime mould transport network restructures in response to a spreading contamination.

\section{Methods}
\label{methods}

\begin{figure}[htbp]
\begin{center}
\subfigure[]{\includegraphics[width=0.8\textwidth]{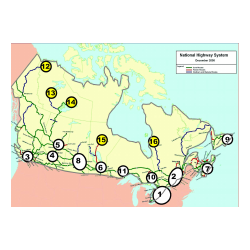}}
\subfigure[]{\includegraphics[width=0.8\textwidth]{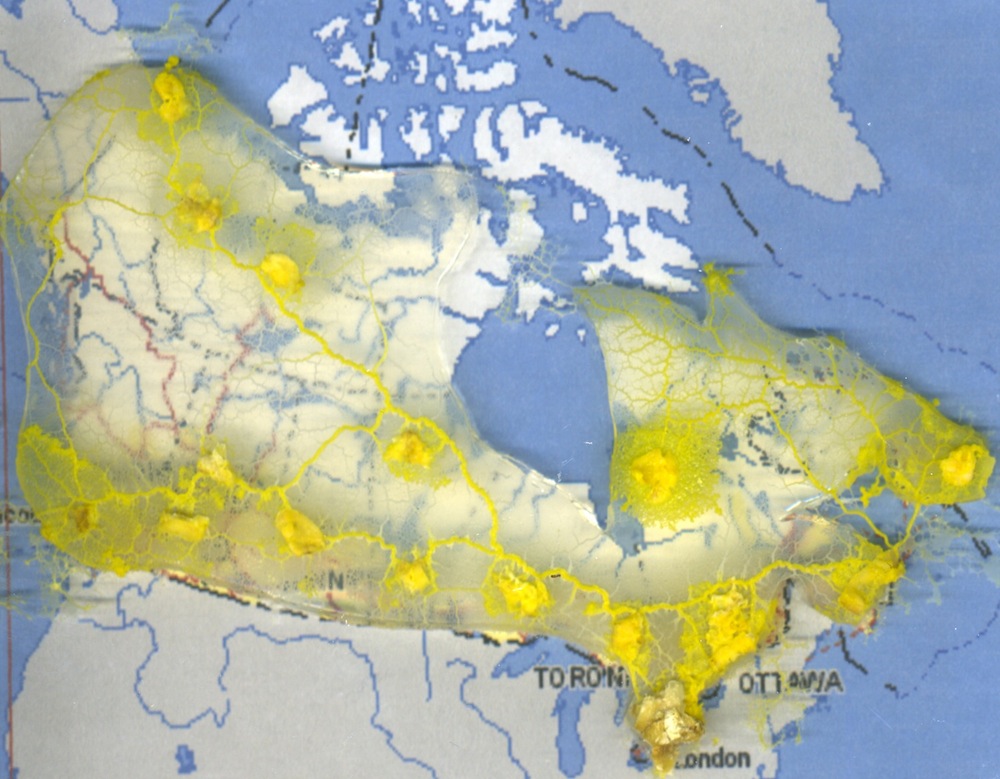}}
\caption{Experimental setup. (a)~Urban areas and transport nodes to be represented by oat flakes,  
from~\cite{CanadaHighways}.
(b)~Snapshot of protoplasmic transport network developed by \emph{P. polycephum}; the snapshot is made on a highway map. }
\label{setup}
\end{center}
\end{figure}

The plasmodium of \emph{ P. polycephalum} is cultivated in plastic containers, on paper kitchen towels sprinkled with
still drinking water and fed with oat flakes (Asda's Smart Price Porridge Oats). For experiments we use
$12 \times 12$~cm polyestyrene square Petri dishes. Agar plates, 2\% agar gel (Select agar, Sigma Aldrich), are 
cut in a shape of Canada. We consider the elven most populated urban areas $\mathbf U$ of Canada (Fig.~\ref{setup}a) and five transport nodes:

\begin{multicols}{2}
\begin{enumerate}
\item Toronto area (including Hamilton,  London,   St. CatharinesÐNiagara,  Windsor,  Oshawa,  
                           Barrie, Guelph, and Kingston)
\item Montreal area (incuding  OttawaÐGatineau, Quebec City, Sherbrooke, Trois-Rivieres) 
\item Vancouver area (including Victoria, Abbotsford, Kelowna)
\item Calgary 
\item Edmonton 
\item Winnipeg 
\item Halifax-Moncton
\item Saskatoon-Regina  
\item St. John's
\item Sudbury
\item Thunder Bay
\item Inuvik
\item Wrigley
\item  Yellowknife
\item  Thompson
\item Radisson
\end{enumerate}
\end{multicols}

The last five entries from Inuvik to Radisson, are not highly populated urban areas. They are transport nodes added for completeness, i.e. to present slime mould with the same number of principle transport nodes as the human-made highways system  (Fig.~\ref{setup}a). Some transport nodes as Fort McMurray, La Ronge, Flin Flon, and so on, are not included in the list due to their proximity to already chosen major urban areas.

To project regions of $\mathbf U$ onto agar gel we place oat flakes in the positions of these regions of  $\mathbf U$ (Fig.~\ref{setup}). In each experiments we tried to match the size and shape of areas in (Fig.~\ref{setup}) by selecting rolled oat of corresponding size and shape. At the beginning of each experiment a piece of plasmodium, usually already attached to an oat flake in the cultivation box, is placed in the Toronto area (region 1 in Fig.~\ref{setup}a). 
The Petri dishes with plasmodium are kept in darkness, at temperature 22-25~C$^{\text o}$, except for observation and image recording.  Periodically (usually in 12~h or 24~h intervals) the dishes are scanned in Epson Perfection 4490. 
We undertook 23 experiments.

\section{Foraging on urban areas}
\label{physarumgraphs}

\begin{figure}[!tbp]
\begin{center}
\subfigure[]{\includegraphics[width=0.4\textwidth]{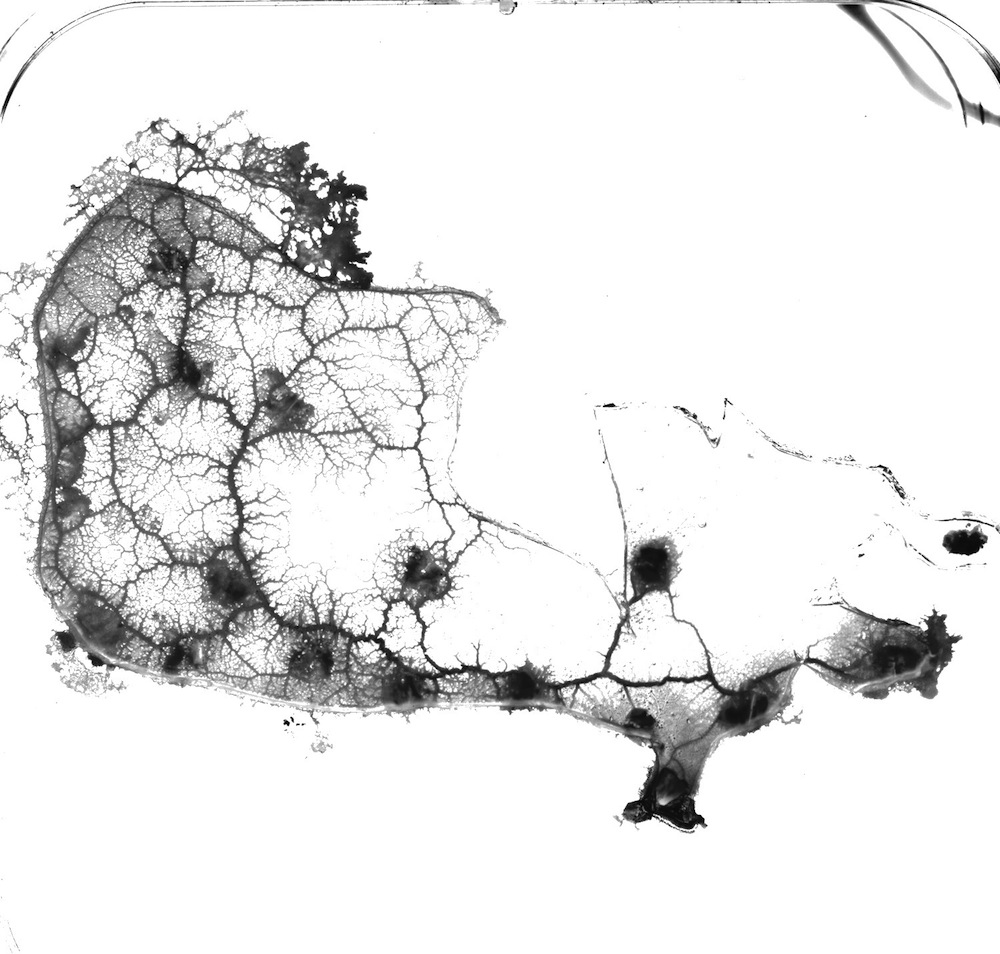}}
\subfigure[]{\includegraphics[width=0.4\textwidth]{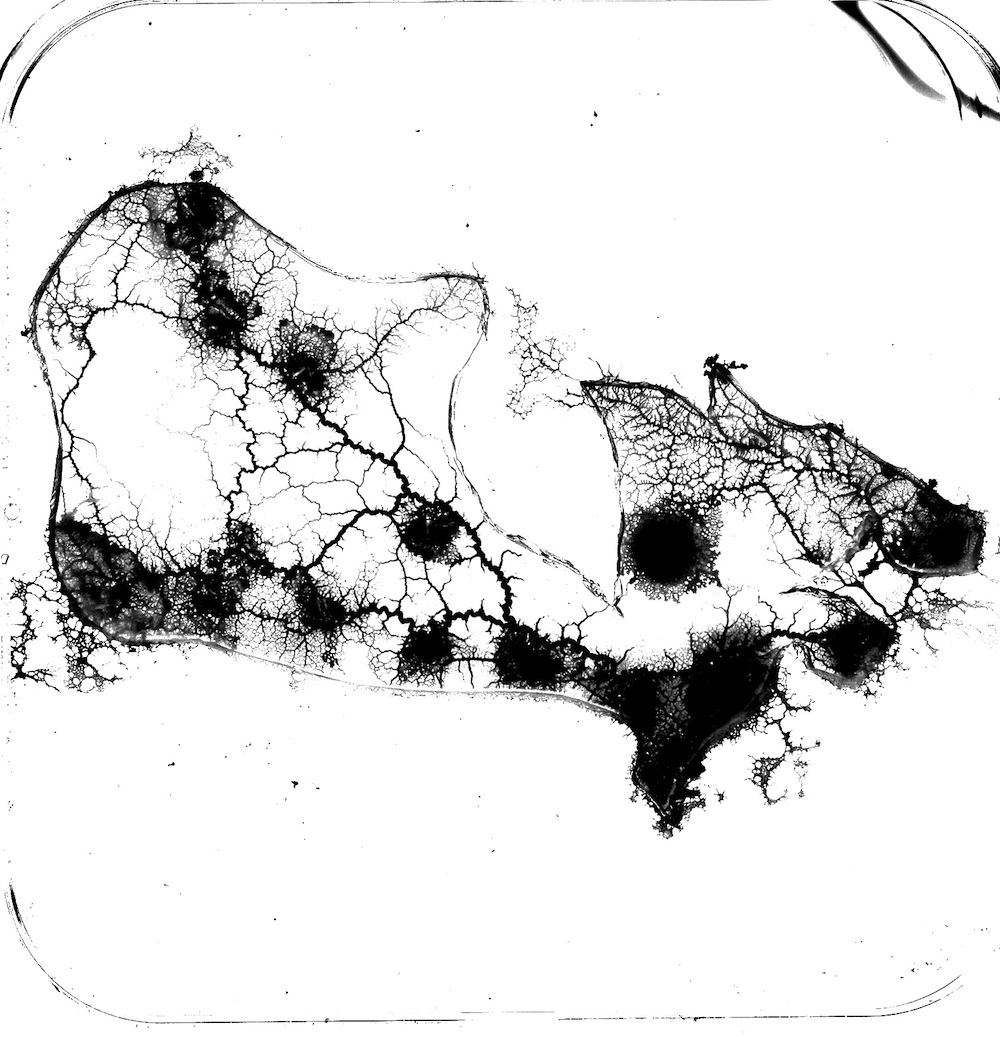}}
\subfigure[]{\includegraphics[width=0.4\textwidth]{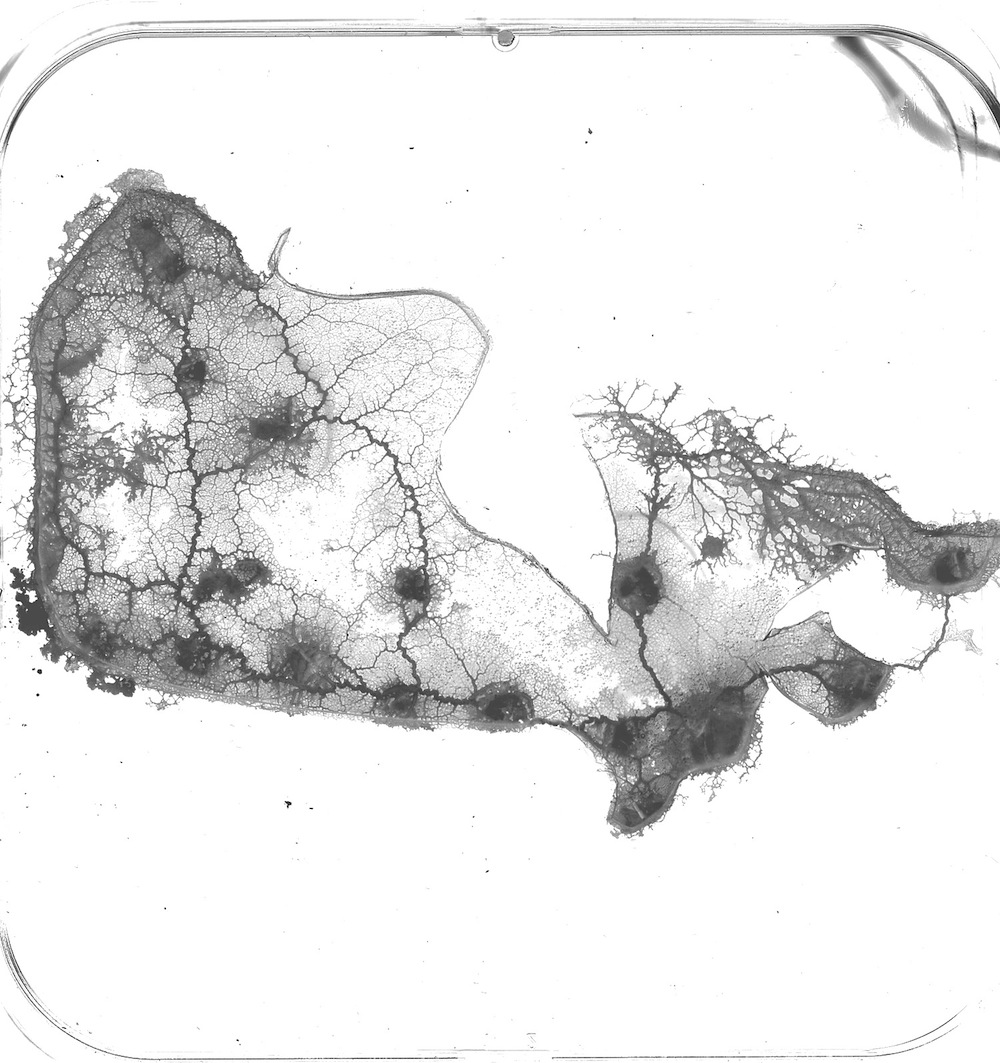}}
\subfigure[]{\includegraphics[width=0.4\textwidth]{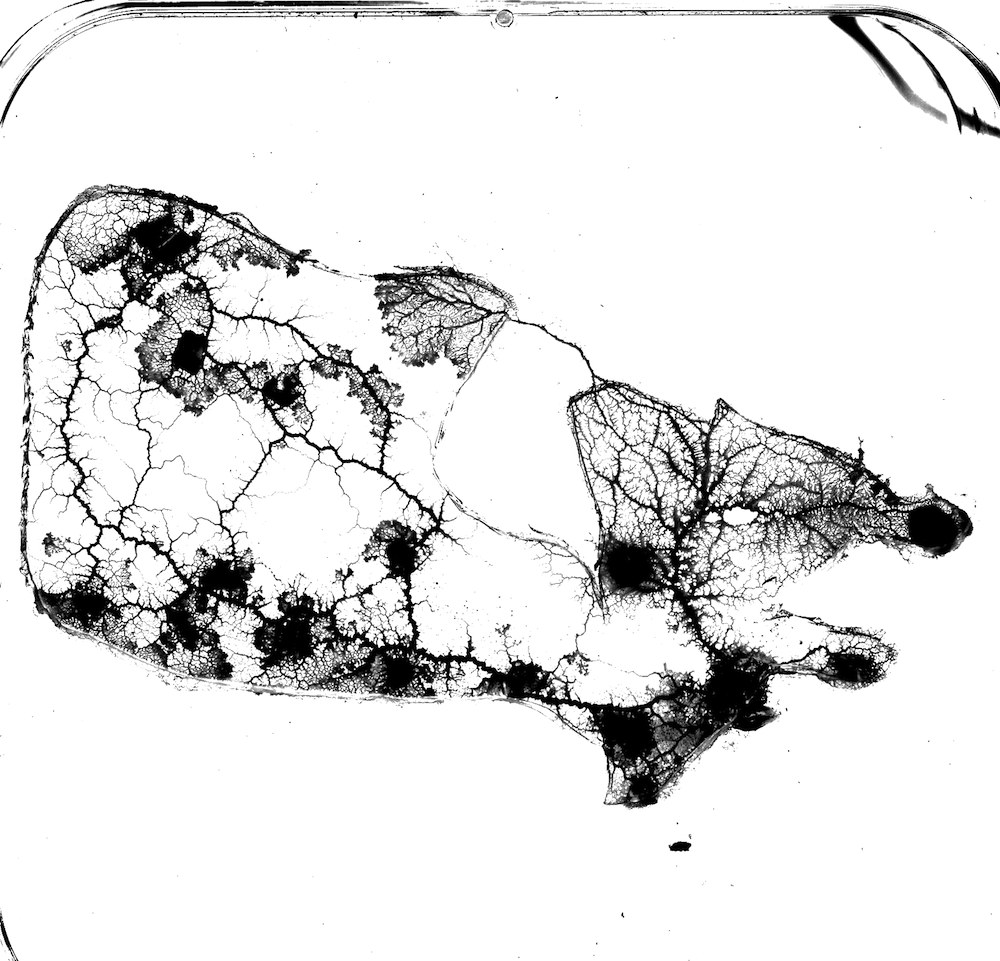}}
\subfigure[]{\includegraphics[width=0.4\textwidth]{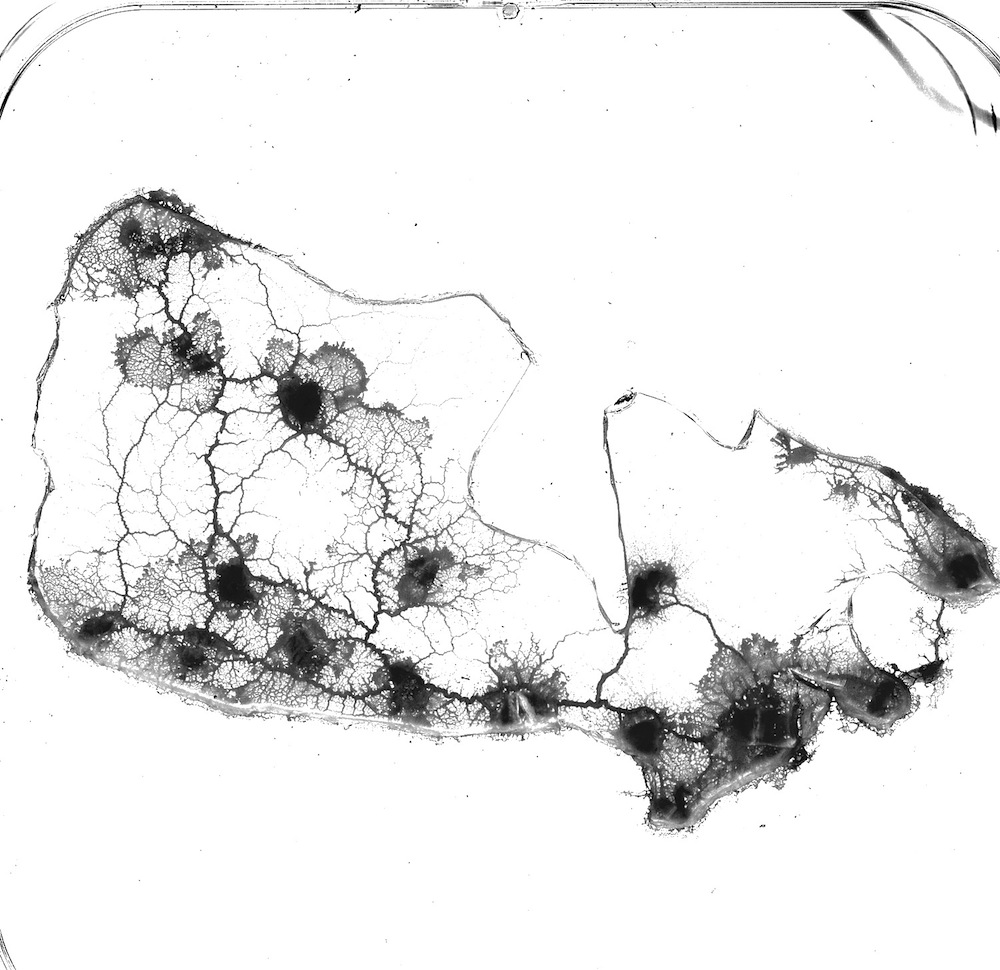}}
\subfigure[]{\includegraphics[width=0.4\textwidth]{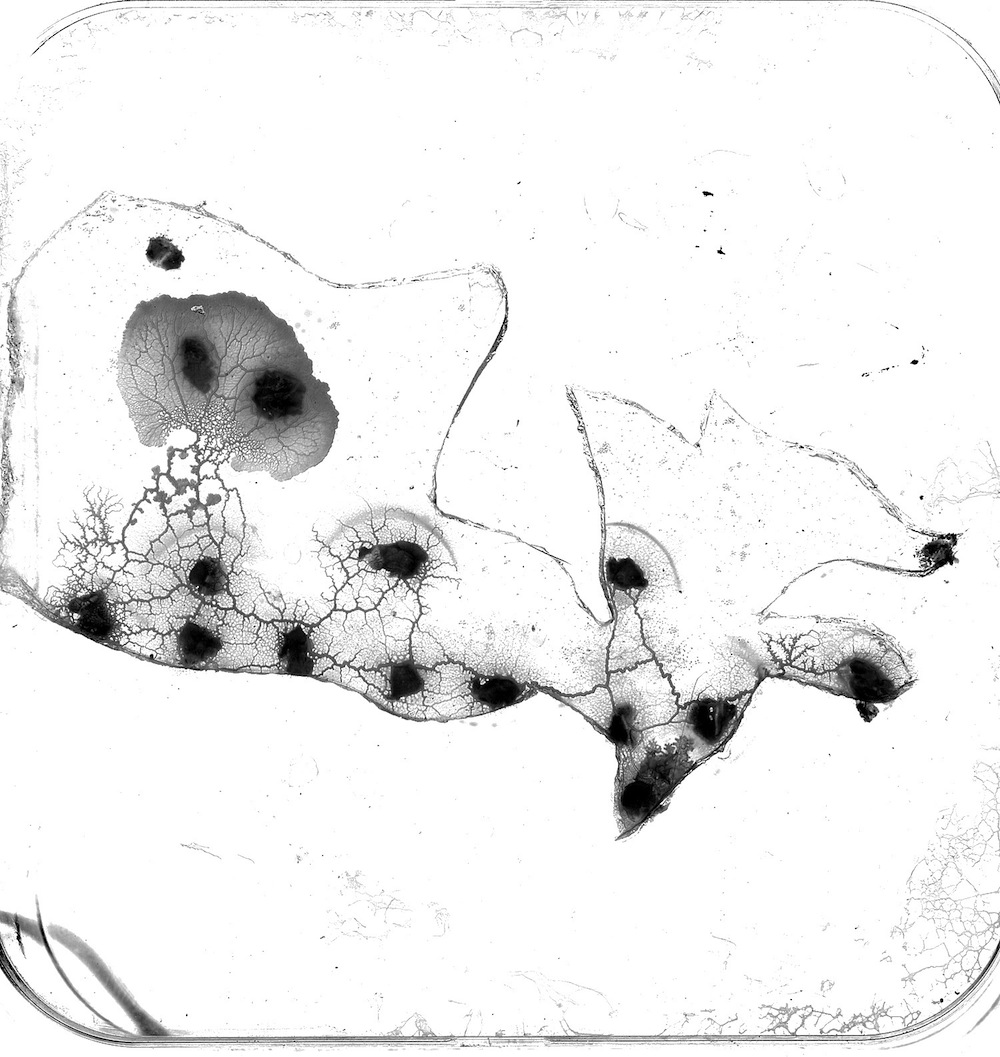}}
\caption{Examples of protoplasmic networks developed by \emph{P. polycephalum} on 
major urban areas and transport nodes $\mathbf U$. Grey-scale images.}
\label{examples}
\end{center}
\end{figure}

It usually takes the plasmodium of \emph{P. polycephalum} 2-5 days to span all urban areas. How fast the plasmodium colonises the space depends on many unknown factors, including seasonal variations, plasmodium's age, etc. 'Younger' plasmodia, which were just recently  'woken up' from the sclerotium phase do usually colonise the experimental arena quicker than old plasmodia, which were replanted several times in culture boxes. Images of protoplasmic networks presented 
in the paper are taken when all oat flakes, representing $\mathbf{U}$, were colonised by plasmodium. Examples
of the protoplasmic networks are shown in Fig.~\ref{examples}.

\begin{figure}[!tbp]
\begin{center}
\subfigure[$\theta=0$]{\includegraphics[width=0.49\textwidth]{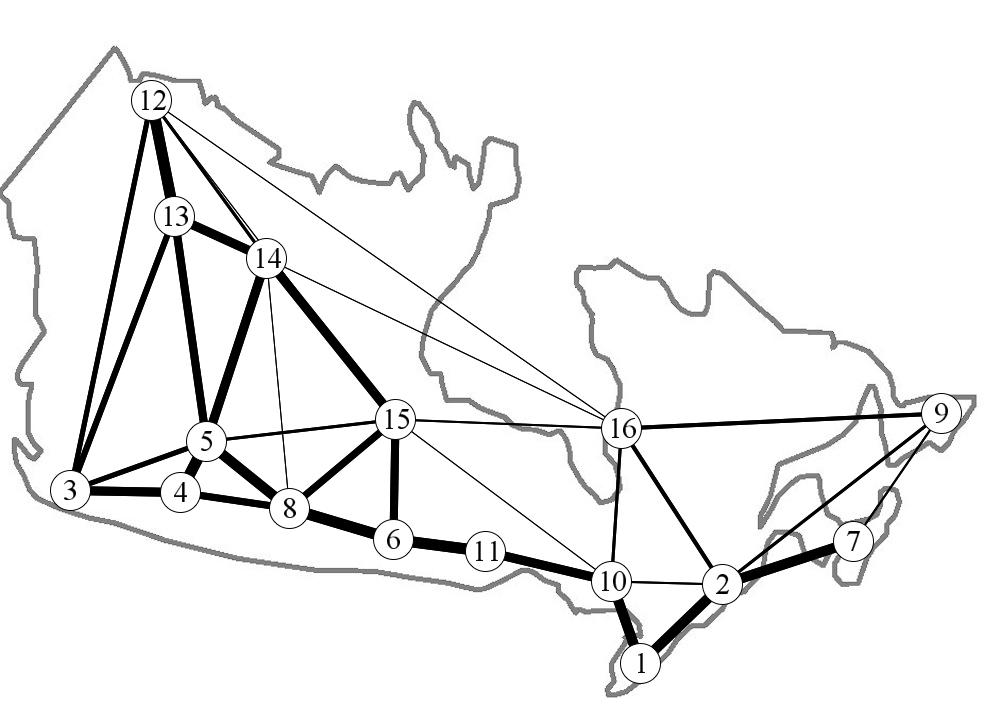}} 
\subfigure[$\theta=\frac{8}{23}$]{\includegraphics[width=0.49\textwidth]{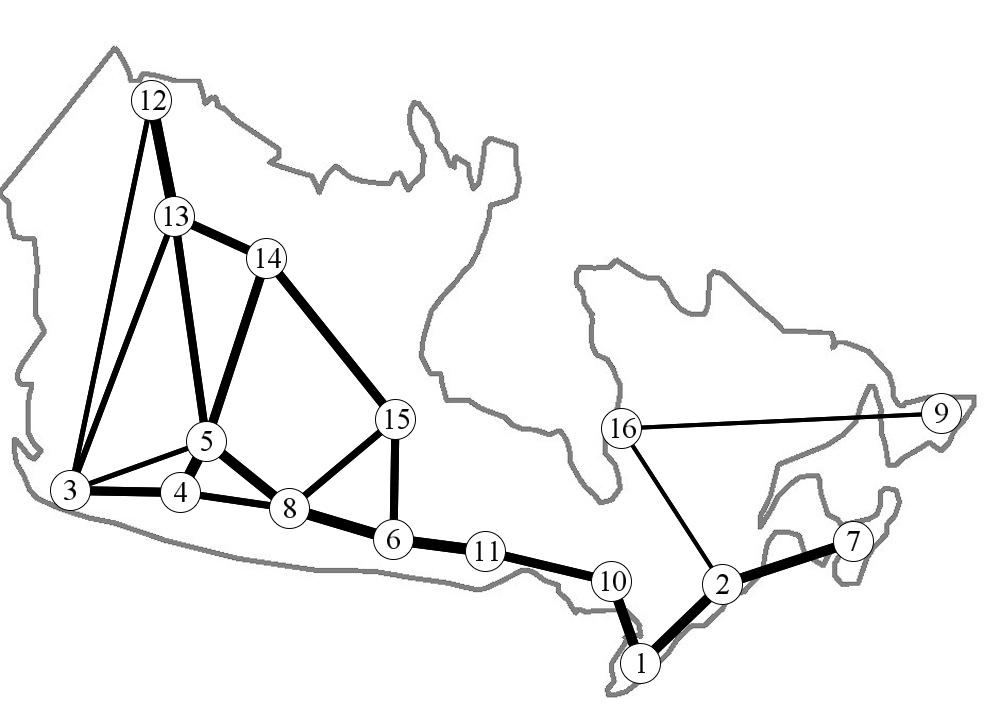}} 
\subfigure[$\theta=\frac{9}{23}$]{\includegraphics[width=0.49\textwidth]{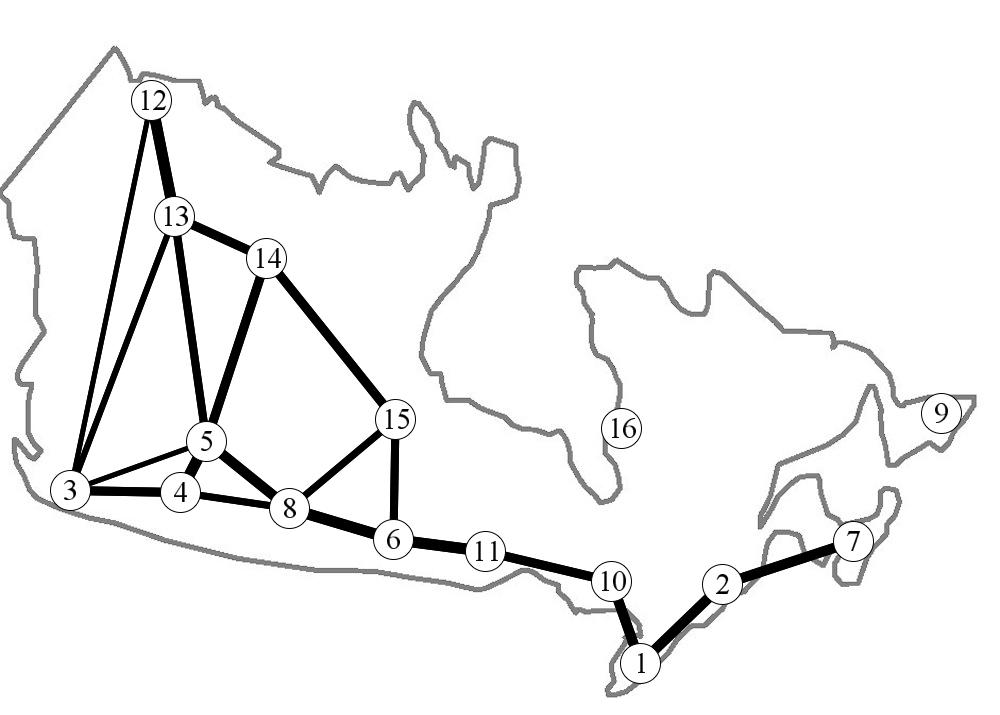}} 
\subfigure[$\theta=\frac{17}{23}$]{\includegraphics[width=0.49\textwidth]{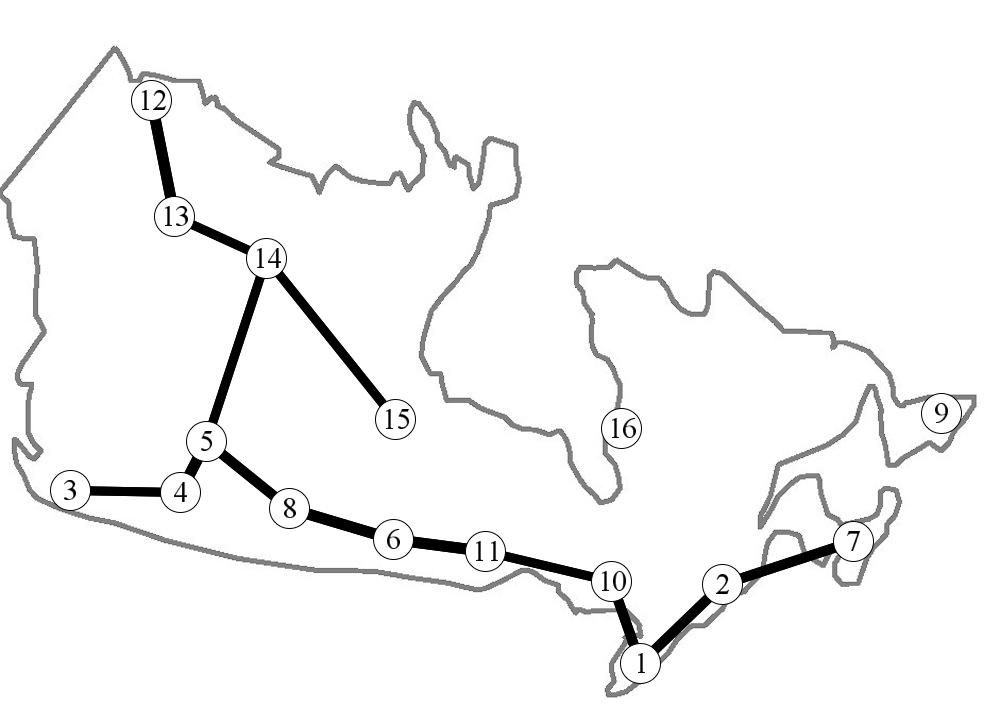}} 
\subfigure[$\theta=\frac{18}{23}$]{\includegraphics[width=0.49\textwidth]{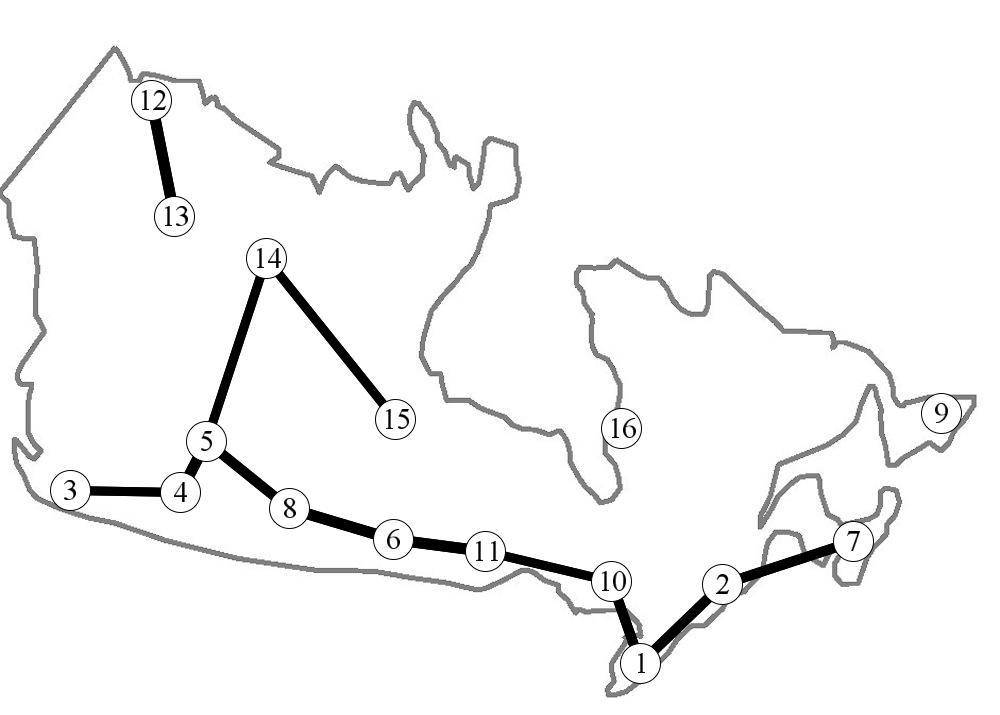}} 
\subfigure[$\theta=\frac{19}{23}$]{\includegraphics[width=0.49\textwidth]{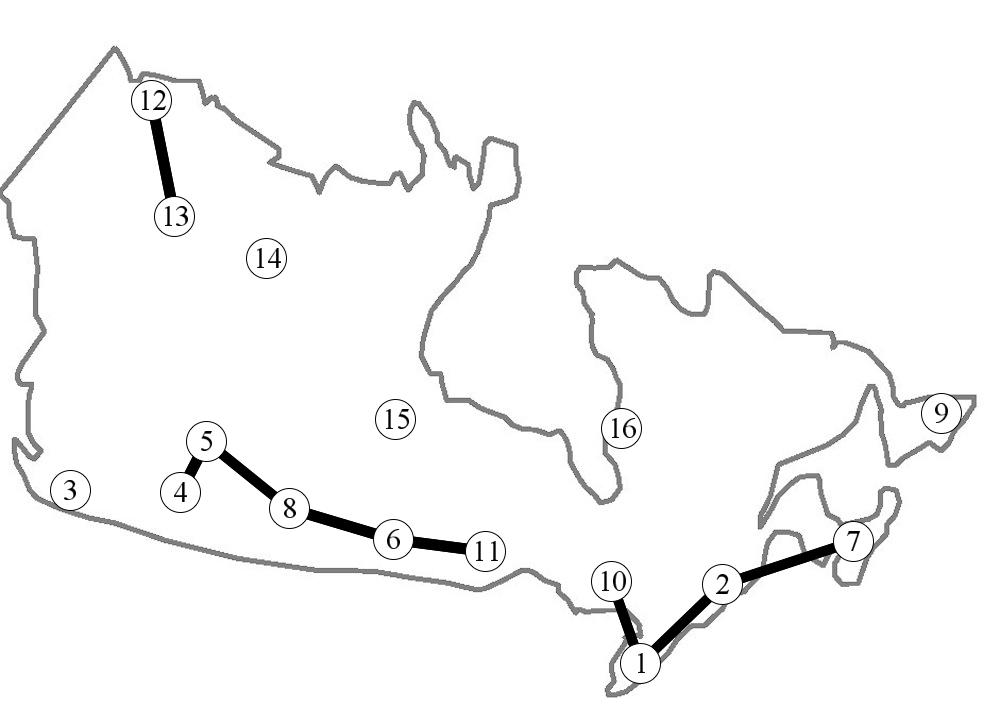}} 
\subfigure[$\theta=\frac{22}{23}$]{\includegraphics[width=0.49\textwidth]{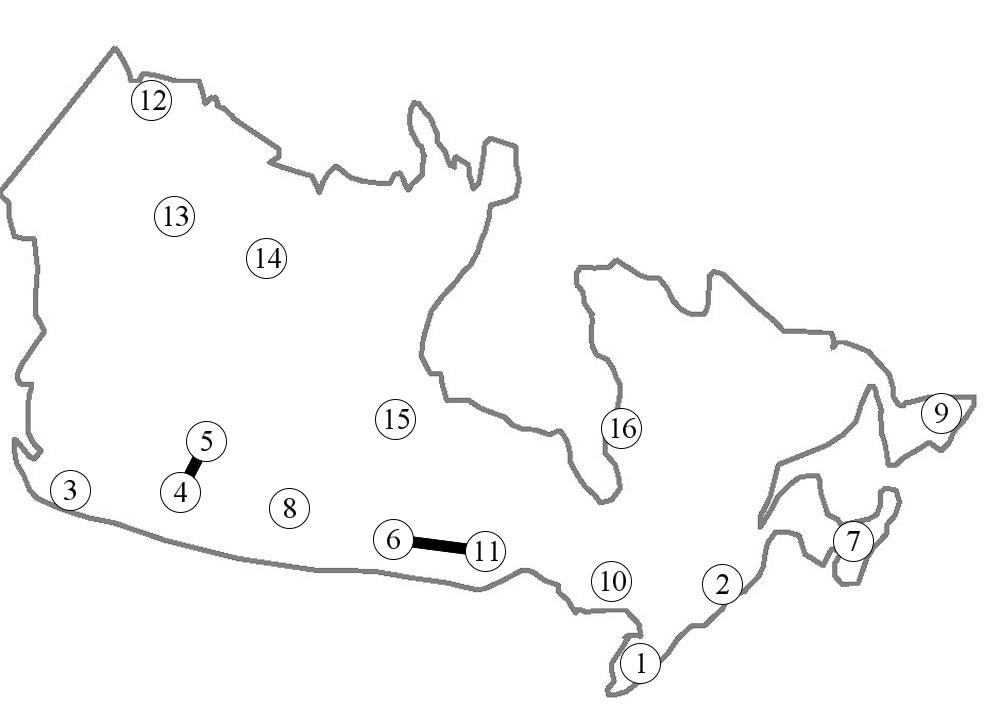}} 
\caption{Configurations \emph{Physarum}-graph ${\mathbf P}(\theta)$ for various cutoff  values of $\theta$.
Thickness of each edge is proportional to the edge's weight.}
\label{physarumgraphs}
\end{center}
\end{figure}

As every living creature does, the plasmodium of \emph{P. polycephalum} rarely repeats its foraging pattern, and almost never builds \emph{exactly} the same protoplasmic network twice. To generalise our experimental results we constructed a Physarum graph with weighted-edges.
A Physarum graph  is a tuple ${\mathbf P} = \langle {\mathbf U}, {\mathbf E}, w  \rangle$, where
$\mathbf U$ is a set of  urban areas,
$\mathbf E$ is a set edges, and
$w: {\mathbf E} \rightarrow [0,1]$ associates each edge of $\mathbf{E}$ with  a probability (or weights).
For every two regions $a$ and $b$ from $\mathbf U$ there is an edge connecting $a$ and $b$ if a plasmodium's protoplasmic link is recorded at least in one of $k$ experiments, and the edge $(a,b)$ has a probability calculated as a ratio of experiments where protoplasmic link $(a,b)$ occurred in the total number of experiments $k=23$. For example, 
if we observed a protoplasmic tube connecting areas $a$ and $b$ in 5 experiments, the weight of edge $(a,b)$ will be $w(a,b)=\frac{5}{23}$. We do not take into account the exact configuration of the protoplasmic tubes but merely their existence. Further we will be dealing with threshold Physarum graphs $\mathbf{P}(\theta)  = \langle  {\mathbf U}, T({\mathbf E}), w, \theta \rangle$. The threshold Physarum graph is obtained from Physarum graph by the transformation:
$T({\mathbf E})=\{ e \in {\mathbf E}: w(e) > \theta \}$. That is all edges with weights less than or equal to $\theta$ are removed.  Examples of threshold Physarum graphs for various values of $\theta$ are shown in 
Fig.~\ref{physarumgraphs}.

A 'raw' Physarum graph $\mathbf{P}(0)$ is a non-planar graph due to the presence of protoplasmic tube connecting Saskatoon-Regina (8) with Yellowknife (14) (Fig.~\ref{physarumgraphs}a). It also exhibits two cross-Canada transport links Inuvik (12) to Radisson (16) and Sudbury (10) to Yellowknife (14). 
Nevertheless, these are links that might be considered as senseless from a geographical point of view
because they are crossing massive of mountains and forests. 

All three links disappear when we increase $\theta$ to 8 (Fig.~\ref{physarumgraphs}b). Four more links become trimmed off when $\theta=8$: Sudbury (10) --- Thompson (15),  Montreal area (2) --- Sudbury (10), Sudbury (10) --- Radisson (16), Montreal area (2) --- St. John's (9).  

$\mathbf{P}(\frac{8}{23})$ is the last connected graph in a series of threshold Physarum graphs, 
$\mathbf{P}(\theta)$, $0 \leq \theta \leq 23$. Urban areas St.~Johns's (9) and Radisson (16) become isolated in 
$\mathbf{P}(\frac{9}{23})$ due to the disappearance of transport links from the Montreal area (2) to Radisson (16) 
and from Radisson  to St. John's (9) (Fig.~\ref{physarumgraphs}c).

The Physarum graph splits into a  tree and two isolated nodes for $\theta=\frac{17}{23}$ (Fig.~\ref{physarumgraphs}d). Five long distance routes from urban areas to transport nodes are removed: 
Vanvouver area (3) -- Inuvik (12), Vancouver area  --- Wrigley (13), 
Edmonton (5) --- Wrigley,  Saskatoon-Regina (8) --- Thompson (15), 
Winnipeg (6) --- Thompson; and, two short-distance routes between major urban areas: 
Vancouver area (3) --- Edmonton (5) and Calgary (4) --- Saskatoon-Regina (8). 

The graph $\mathbf{P}(\frac{18}{23})$ consists of several disconnected components: two isolated nodes, St.~John's 
and Sudbury, 
 two-node segment Inuvik --- Wrigley, and a tree spanning the rest of the urban ares (Fig.~\ref {physarumgraphs}e). A further increase of 
$\theta$ to $\frac{19}{23}$ (Fig.~\ref {physarumgraphs}f) leads to the formation of 
\begin{itemize}
\item five isolated nodes: Vancouver area, St.~John's, Yellowknife, Thompson, Winnipeg; 
\item a segment: Inuvik --- Wrigley;
\item  a chain: Calgary --- Edmonton --- Saskatoon-Regina --- Winnipeg --- Thunder Bay; 
\item a chain: Sudbury --- Toronto area --- Montreal area --- Halifax-Moncton. 
\end{itemize}

Only  segment-routes Calgary to Edmonton and Winnipeg to Thunder Bay  are represented by protoplasmic tubes in almost all experiments  (Fig.~\ref {physarumgraphs}g).

\section{Physarum network vs highway network}
\label{PhysarumVsHighways}

\begin{figure}[!tbp]
\begin{center}
\subfigure[]{\includegraphics[width=0.49\textwidth]{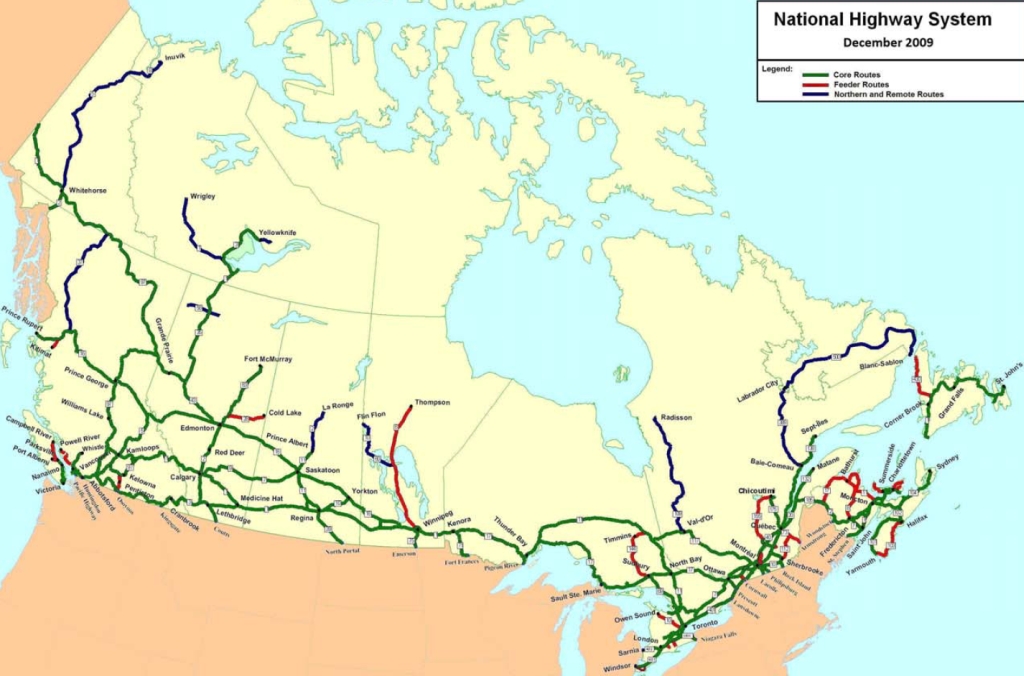}}
\subfigure[$\mathbf{H}$]{\includegraphics[width=0.49\textwidth]{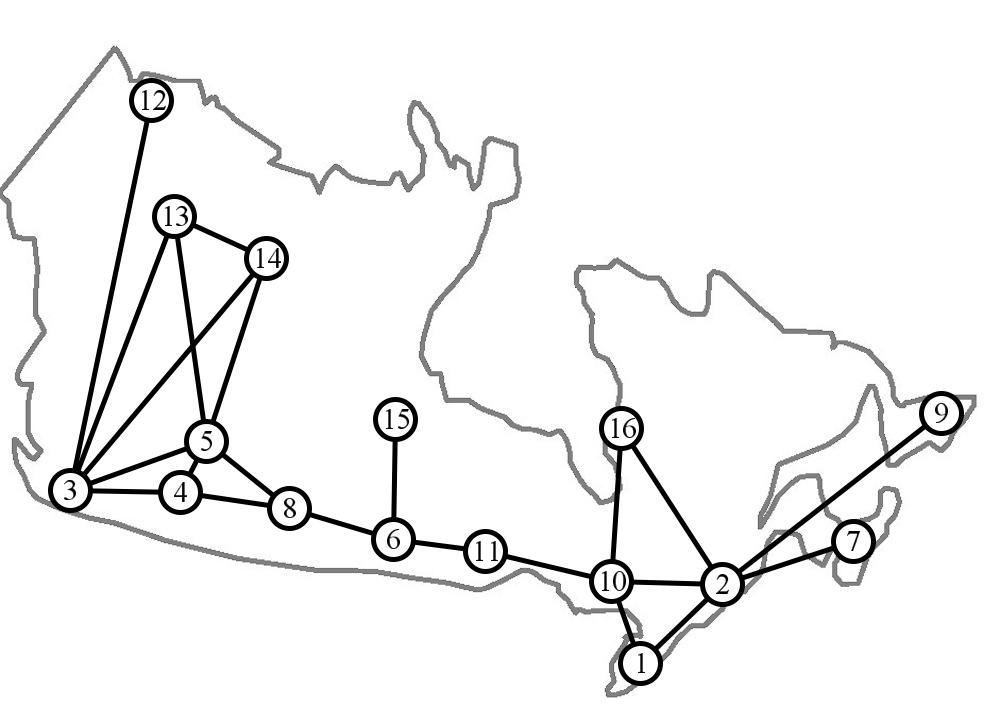}}
\subfigure[$\mathbf{P}(0) \bigcap {\mathbf H}$]{\includegraphics[width=0.49\textwidth]{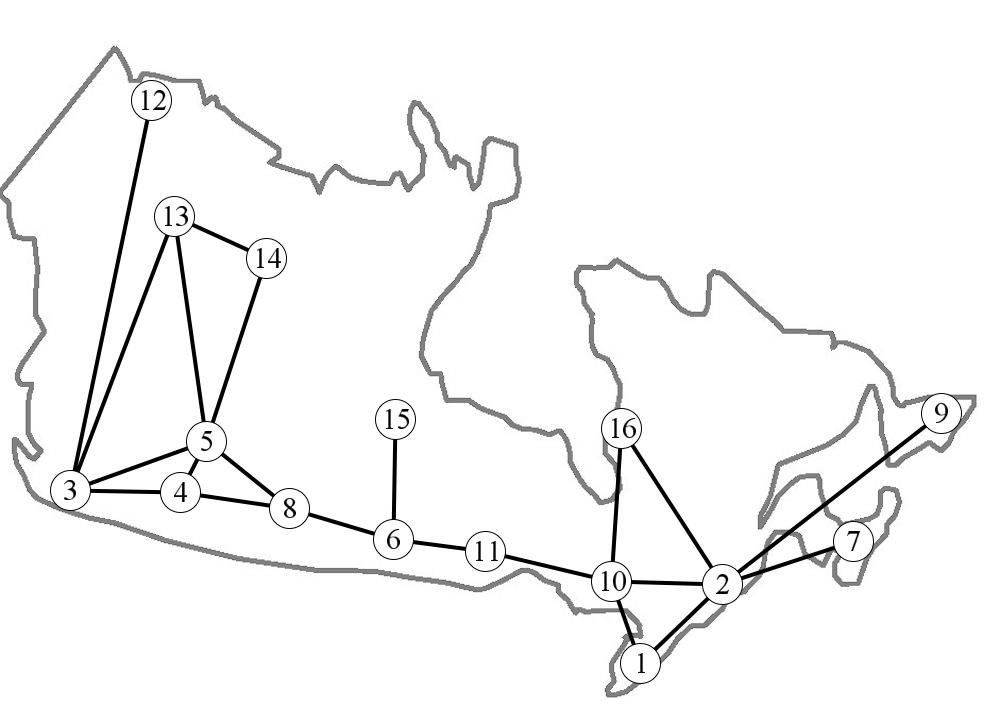}}
\subfigure[$\mathbf{P}(\frac{8}{23}) \bigcap {\mathbf H}$]{\includegraphics[width=0.49\textwidth]{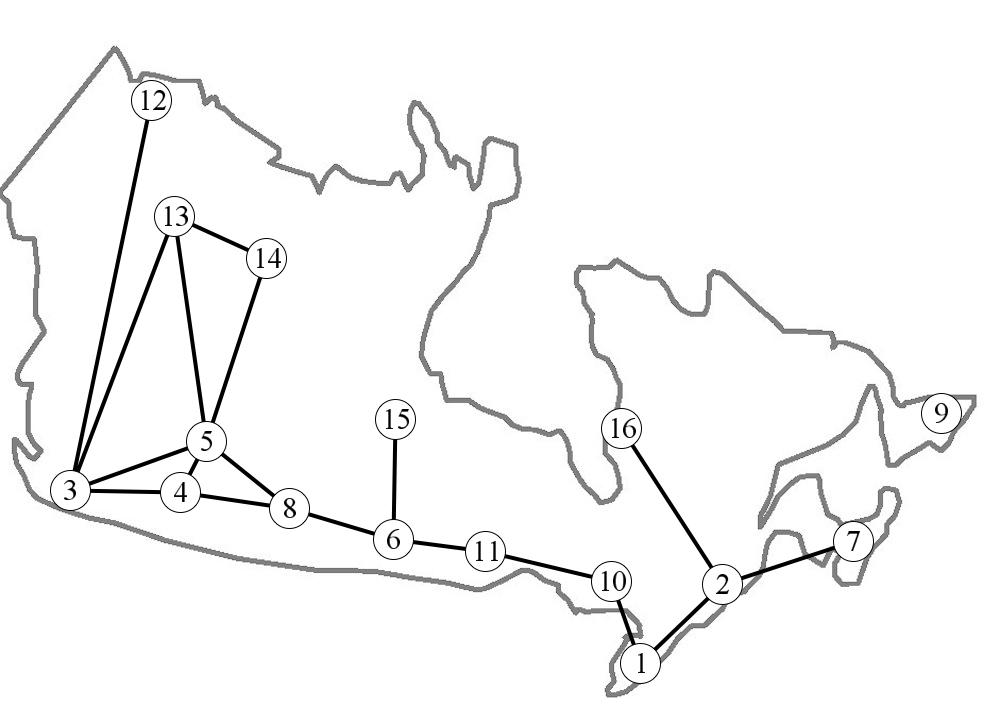}}
\subfigure[$\mathbf{P}(\frac{17}{23}) \bigcap {\mathbf H}$]{\includegraphics[width=0.49\textwidth]{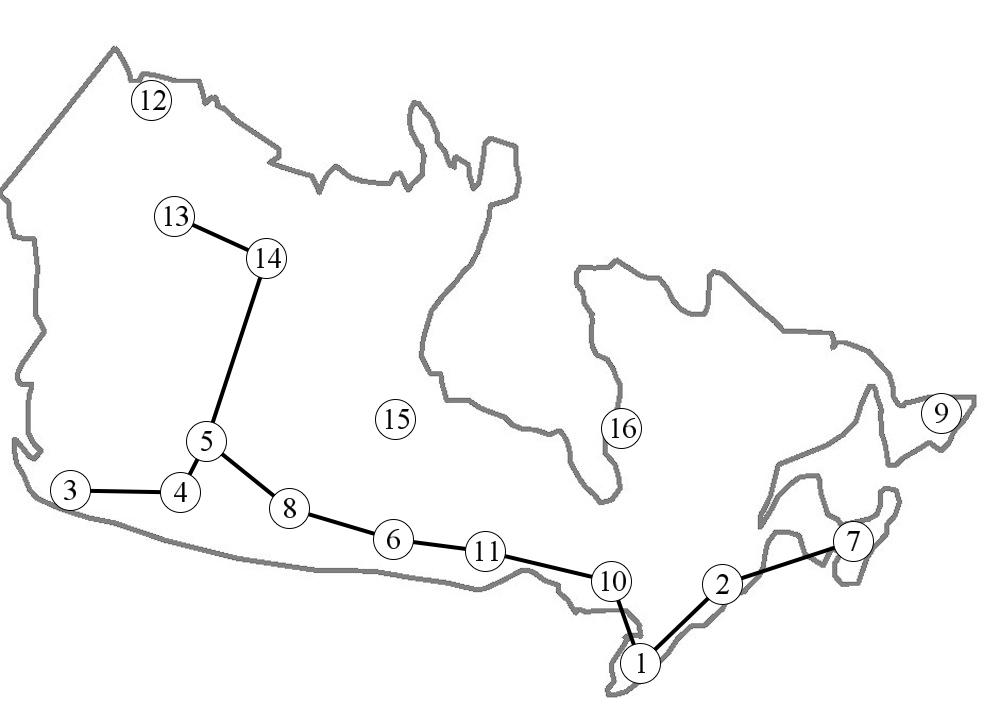}}
\caption{Physarum vs highway network. (a)~Highway network in Canada~\cite{CanadaHighways}.
(b)~Highway graph $\mathbf H$. (cde)~Intersection of threshold Physarum graph with highway graph for 
$\theta=0, \frac{8}{23}, \frac{17}{23}$.}
\label{highways}
\end{center}
\end{figure}

 We construct the highway graph $\mathbf H$ as follows. Let $\mathbf U$ be a set of urban regions, for any two regions $a$ and $b$ from $\mathbf U$, the nodes $a$ and $b$ are connected by an edge ($a$, $b$) if there is a motorway starting in the vicinity of $a$ and passing in the vicinity of $b$ and not passing in the vicinity of any other urban area $c \in \mathbf U$. Highway graph $\mathbf{H}$ shown in Fig.~\ref{highways}b is extracted from a scheme of the Canadian transport network (Fig.~\ref{highways}a).

in the vicinity of a and passing in the vicinity of b and not passing in the vicinity

\begin{finding}
Physarum almost approximates the Canadian highway network.
\end{finding}

'Raw' Physarum graph $\mathbf{P}(0)$ approximates 21 of 22 edges of highway graph $\mathbf{H}$.
Only one edge, Vancouver area to Calgary, of the highway graph $\mathbf{H}$ is not represented 
by protoplasmic tubes in any of the 23 experiments undertaken (Fig.~\ref{highways}c). 
The 'raw' physarum graph gives us a rather relaxed approximation because it includes even links which occurred just once in a set of experiments.
Let us look at 
$\mathbf{P}(\frac{8}{23})$ which represents links which occurred in over 35\% of experiments.  Physarum graph 
$\mathbf{P}(\frac{8}{23})$ approximates 18 of 22 edges of $\mathbf{H}$ (Fig.~\ref{highways}d). The only edges
of $\mathbf{H}$ not represented in $\mathbf{P}(\frac{8}{23})$ are Vancouver area to Yellowknife,  Sudbury to Radisson, Sudbury to Montreal area, and St. John's to Montreal area.

\begin{finding}
A core component of the Physarum transport network and the Canadian highway network consists of  a chain passing along the south border from Halifax-Monctron area to Edmonton, and a fork attached to Edmonton; the south branch of the fork 
is Edmonton --- Calgary --- Vancouver area and the  north branch is Edmonton -- Yellowknife -- Wrigley. 
\end{finding}

The component above is the only connected component in the intersection of 
$\mathbf{P}(\frac{17}{23})$ with $\mathbf{H}$ (Fig.~\ref{highways}e).

\section{Proximity graphs}
\label{proximitygraphs}

\begin{figure}[!tbp]
\begin{center}
\subfigure[$\mathbf{GG}$]{\includegraphics[width=0.49\textwidth]{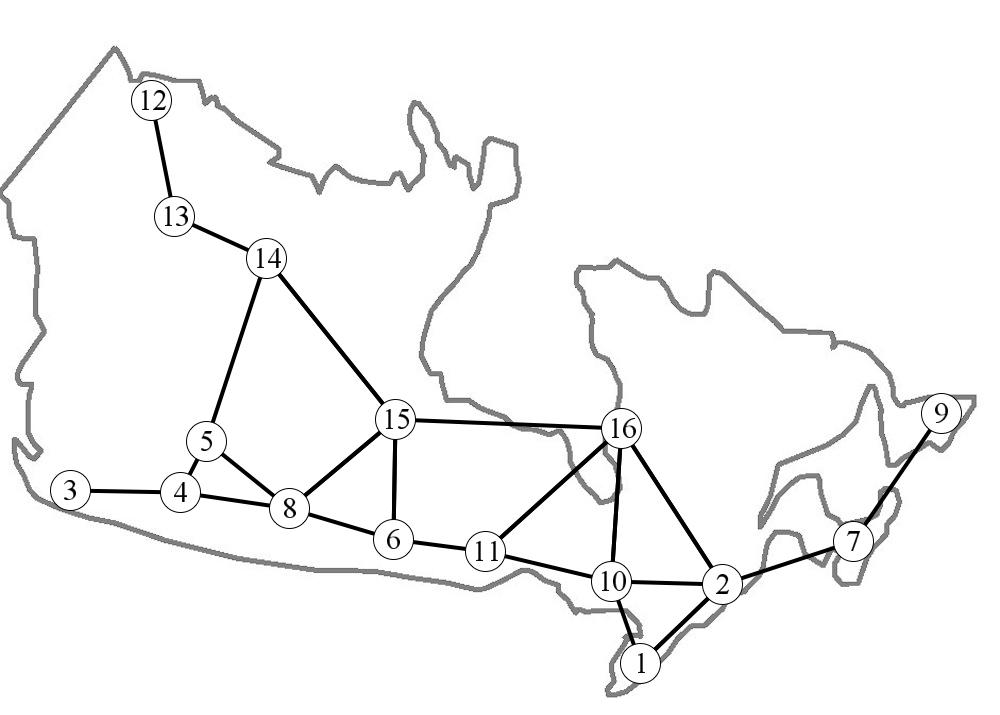}}
\subfigure[$\mathbf{RNG}=\mathbf{MST}$]{\includegraphics[width=0.49\textwidth]{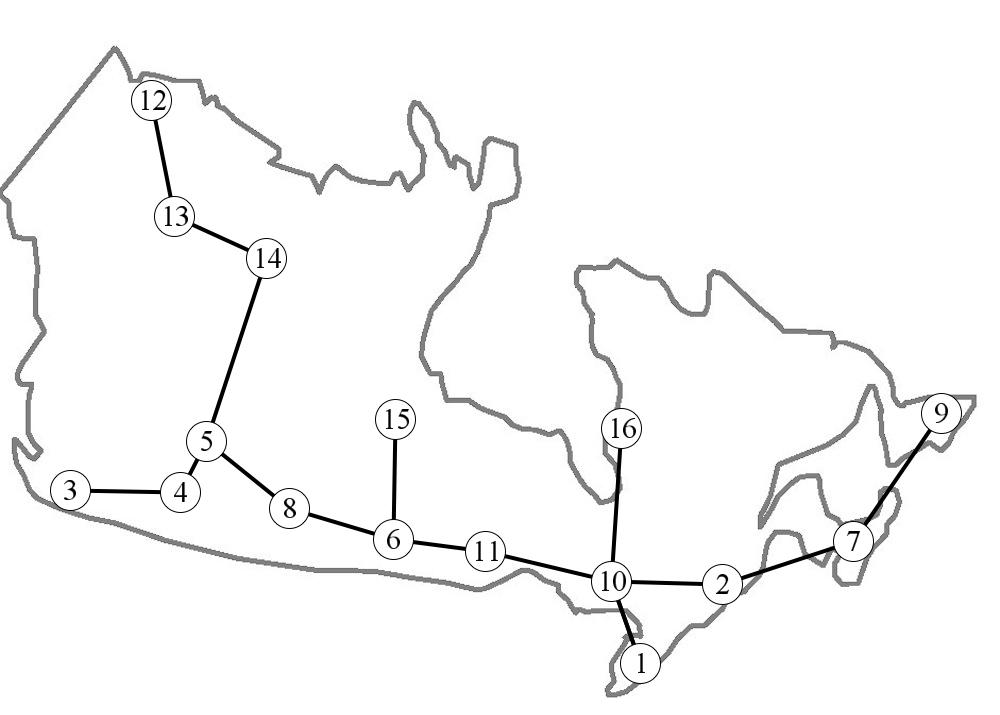}}
\subfigure[$\mathbf{GG} \bigcap {\mathbf H}$]{\includegraphics[width=0.49\textwidth]{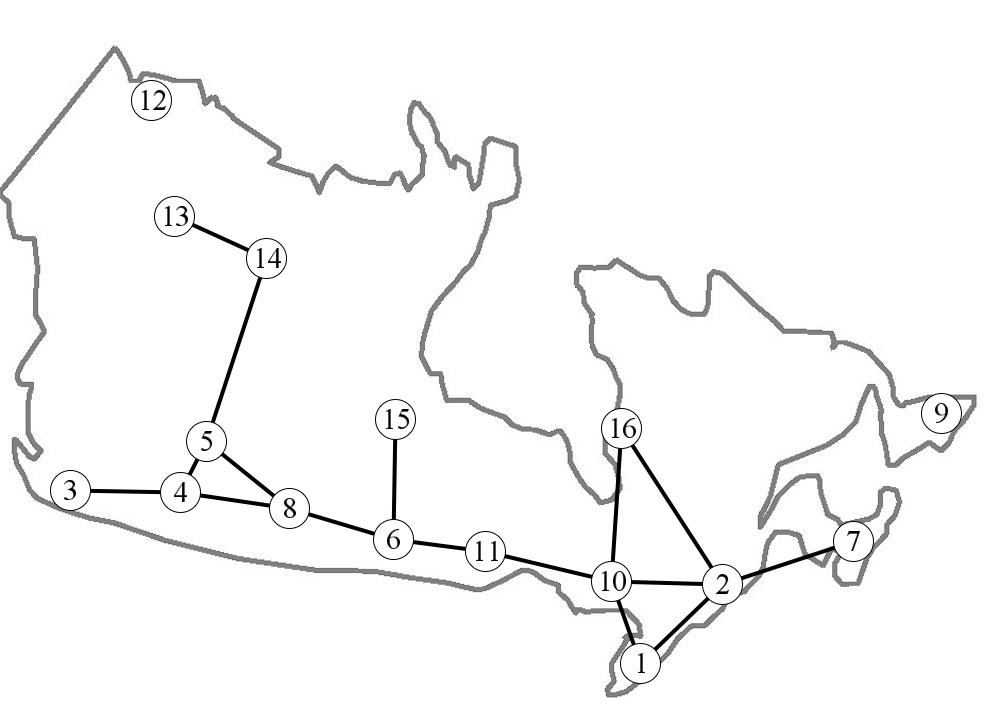}}
\subfigure[$\mathbf{RNG} \bigcap {\mathbf H}$]{\includegraphics[width=0.49\textwidth]{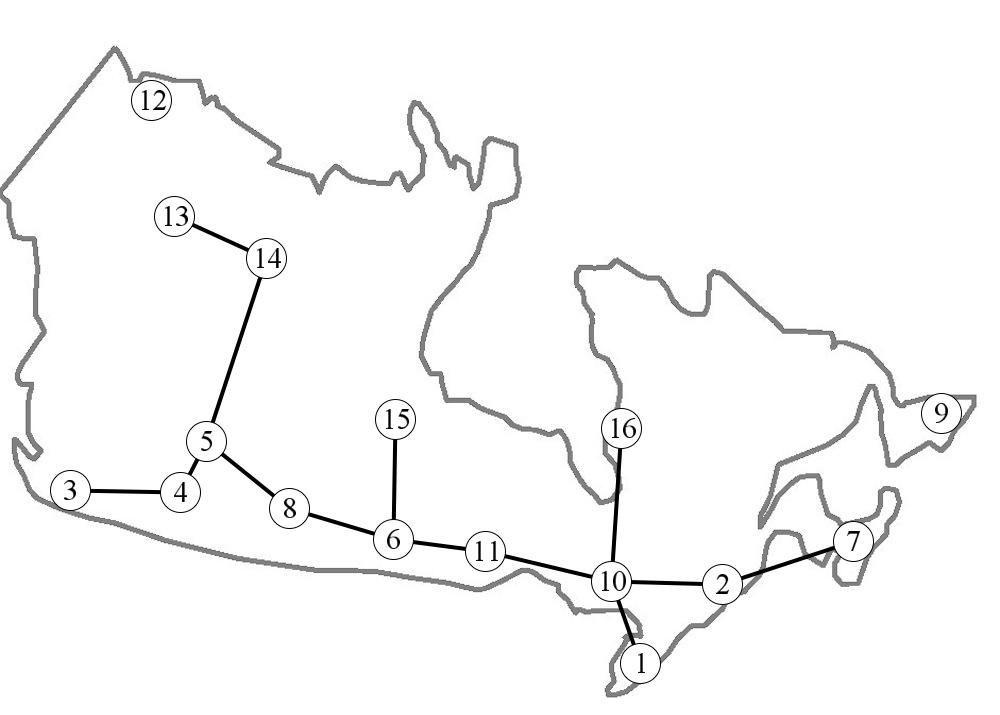}}
\caption{Proximity graphs and their intersection with highway graph $\mathbf H$. (a)~Gabriel graph.
(b)~Relative neighbourhood graph and minimun spanning tree. (c)~Intersection of Gabriel graph with highway graph. 
(d)~Intersection of relative neighbourhood graph with highway graph.}
\label{proximity}
\end{center}
\end{figure}

A planar graph consists of nodes which are points of the  Euclidean plane and edges which are straight segments connecting the points. A planar proximity graph is a planar graph where two points are connected by an edge if they are close in some sense. A pair of points is assigned a certain neighbourhood, and points of the pair are connected by an edge if their neighbourhood is empty.  Here we consider the most common proximity graph as follows.
\begin{itemize}
\item $\mathbf{GG}$: Points $a$ and $b$ are connected by an edge in the Gabriel Graph $\mathbf{GG}$ if
disc with diameter $dist(a,b)$ centered in middle of the segment $ab$ is empty~\cite{gabriel_sokal_1969, matula_sokal_1984} (Fig.~\ref{proximity}a).
\item $\mathbf{RNG}$: Points $a$ and $b$ are connected by an edge in the Relative Neighbourhood Graph $\mathbf{RNG}$ if no other point $c$ is closer
to $a$ and $b$ than $dist(a,b)$~\cite{toussaint_1980} (Fig.~\ref{proximity}b).
\item $\mathbf{MST}$: The Euclidean minimum spanning tree (MST)~\cite{nesetril} is a connected acyclic graph which has minimum possible sum of edges' lengths (Fig.~\ref{proximity}b).
\end{itemize}
In general, the graphs relate as
$\mathbf{MST} \subseteq \mathbf{RNG}  \subseteq\mathbf{GG}$~\cite{toussaint_1980,matula_sokal_1984,jaromczyk_toussaint_1992}.

\begin{finding}
For a given configuration of nodes of $\mathbf{U}$ $\mathbf{RNG} = \mathbf{MST}$.
\end{finding}

The finding implies that the configuration of urban areas of $\mathbf{U}$ is 'spanning friendly'. 

\begin{finding}
Let $\mathbf{T}_i$ be a minimum spanning tree rooted in node $i \in \mathbf{U}$ then 
$\mathbf{T}_i = \mathbf{T}_j$ for any $i,j \in \mathbf{U}$.
\end{finding}

We demonstrated this by direct computation of all possible spanning trees on $\mathbf{U}$.

\begin{finding}
St. John's and Inuvik urban areas are isolated in $\mathbf{GG}  \bigcap \mathbf{RNG}$ and 
$\mathbf{GG}  \bigcap \mathbf{MST}$.
\end{finding}

\begin{finding}
$\mathbf{MST}$/\{(Inuvik --- Wrigley), (Halifax-Moncton --- St. John's)\} $\subset$ $\mathbf{H}$  
\end{finding}

This means that Canadian highway network is almost optimal (Fig.~\ref{proximity}de). 

\begin{figure}[!tbp]
\begin{center}
\subfigure[$\mathbf{P}(\frac{0}{23}) \bigcap {\mathbf{GG}}$]
{\includegraphics[width=0.49\textwidth]{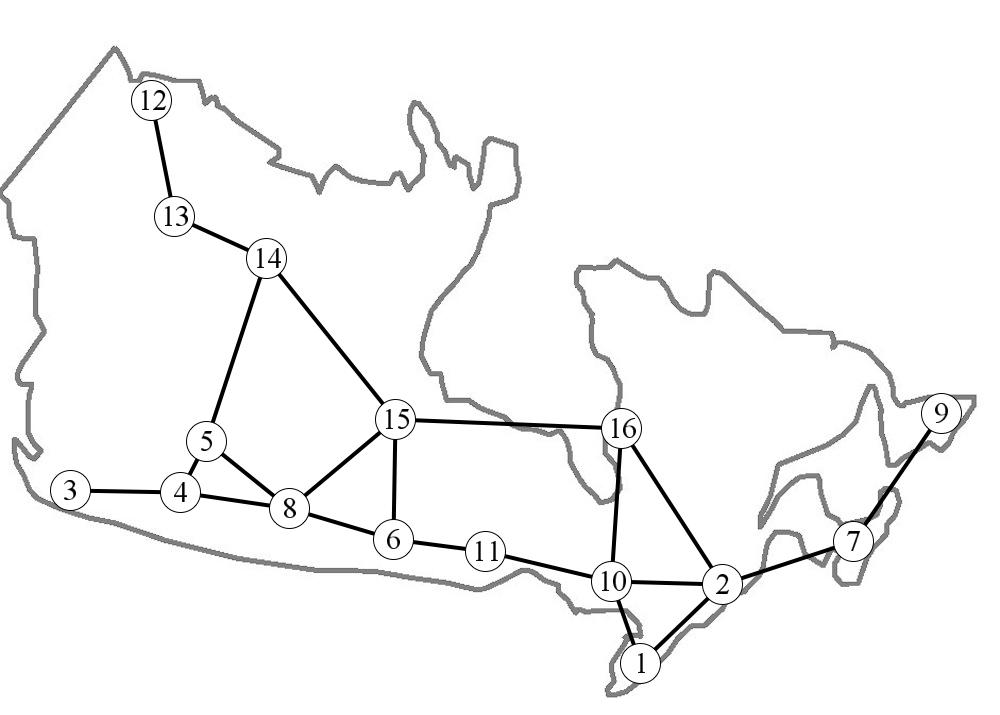}}
\subfigure[$\mathbf{P}(\frac{0}{23}) \bigcap {\mathbf{MST}}$]
{\includegraphics[width=0.49\textwidth]{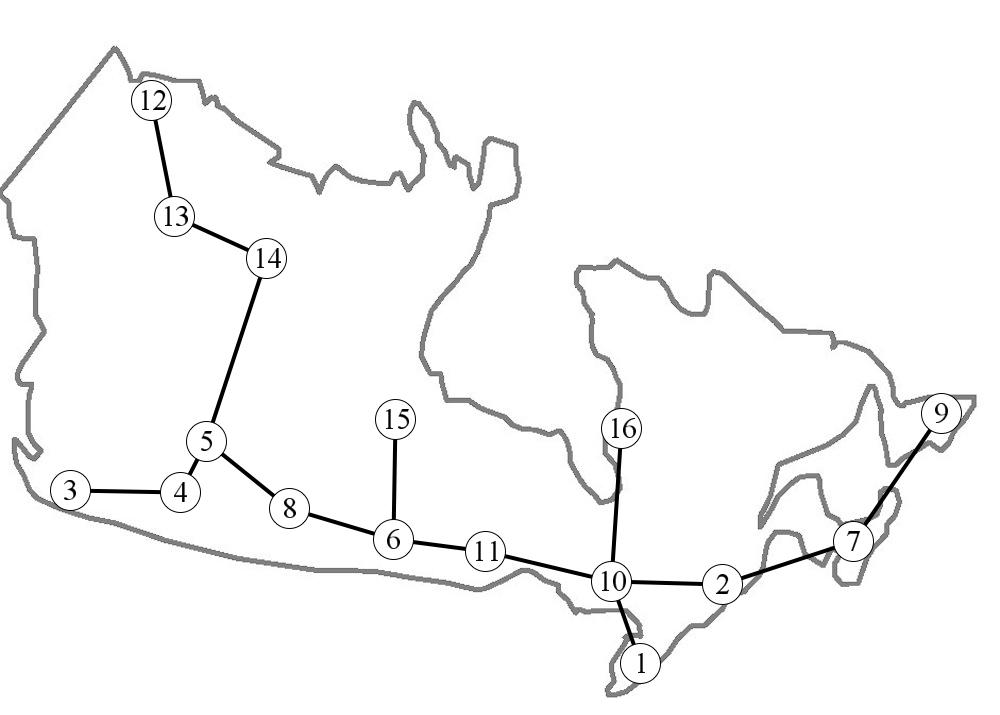}}
\subfigure[$\mathbf{P}(\frac{8}{23}) \bigcap {\mathbf{GG}}$]
{\includegraphics[width=0.49\textwidth]{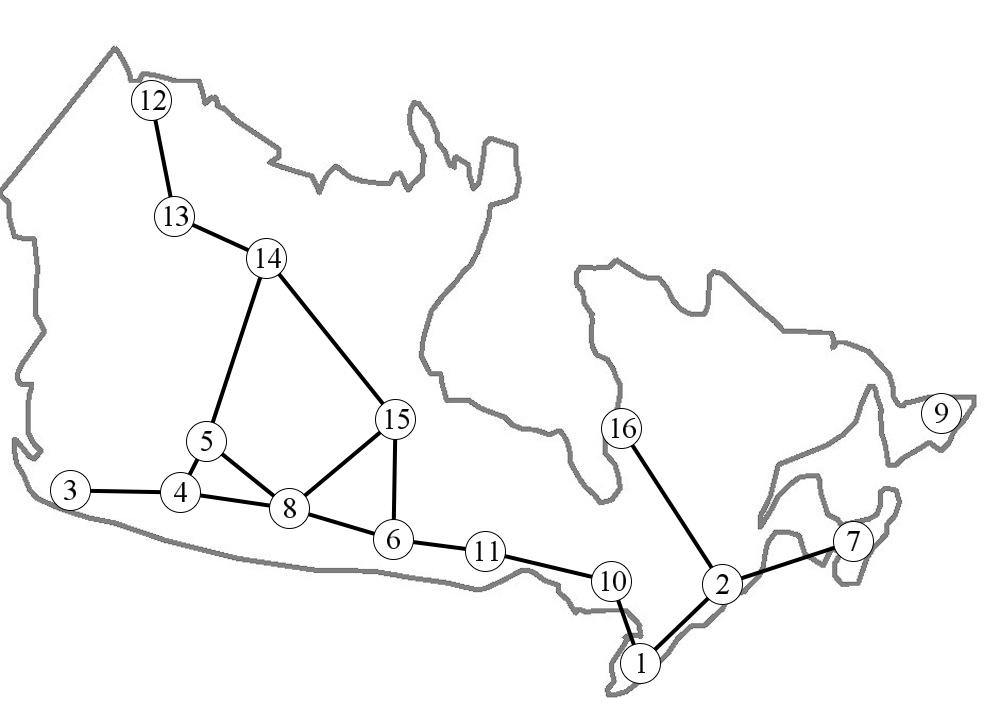}}
\subfigure[$\mathbf{P}(\frac{8}{23}) \bigcap {\mathbf{MST}}$]
{\includegraphics[width=0.49\textwidth]{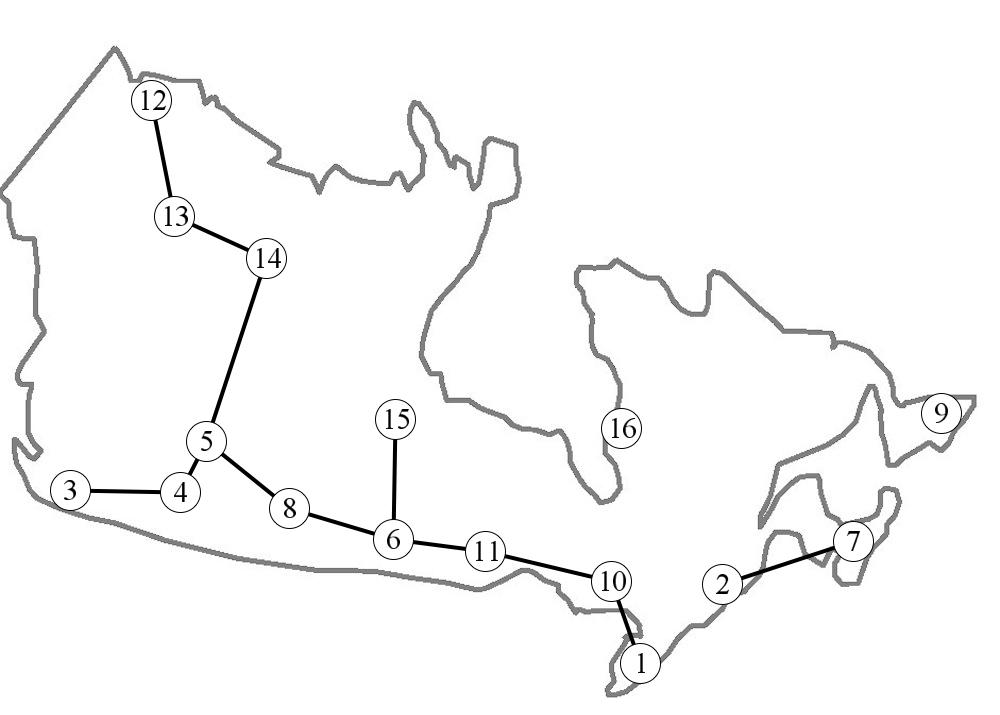}}
\subfigure[$\mathbf{P}(\frac{17}{23}) \bigcap {\mathbf{GG}}$]
{\includegraphics[width=0.49\textwidth]{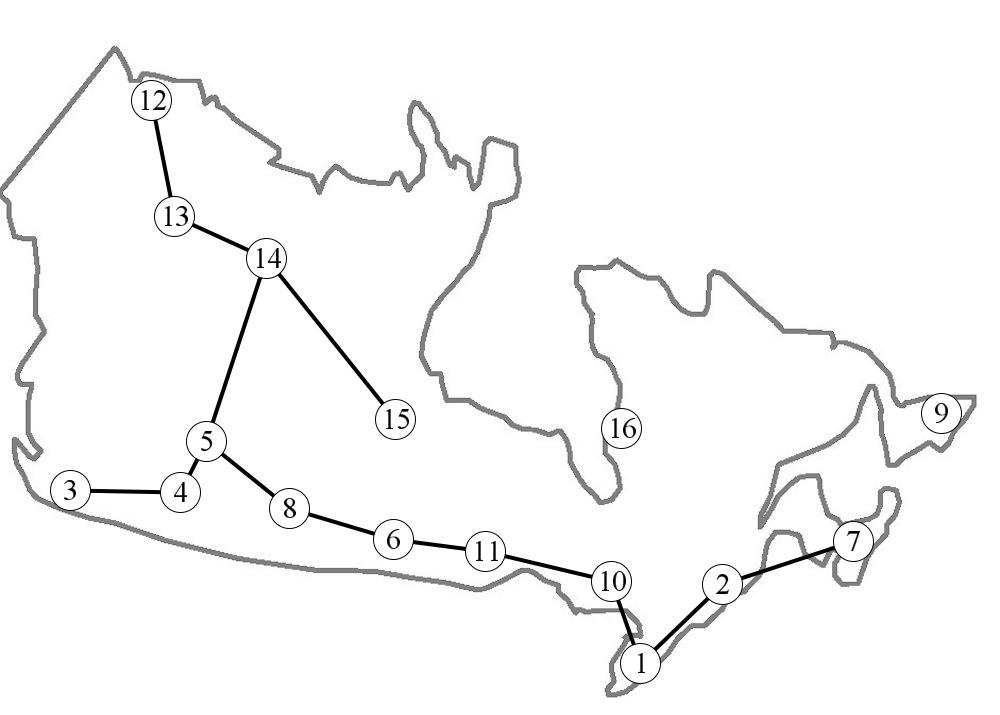}}
\subfigure[$\mathbf{P}(\frac{17}{23}) \bigcap {\mathbf{MST}}$]
{\includegraphics[width=0.49\textwidth]{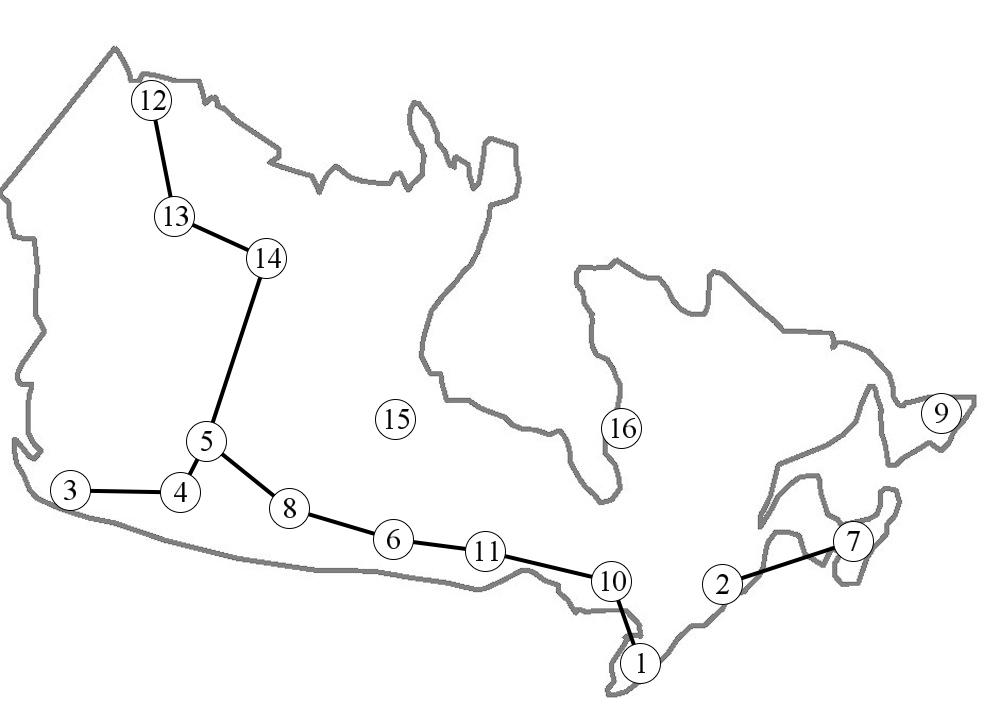}}
\caption{Intersection of (ace)~Gabriel graph $\mathbf{GG}$  
and (bdf)~minimum spanning tree $\mathbf{MST}$ with Physarum graphs $\mathbf{P}(\theta)$
for (ab)~$\theta=0$, (cd)~$\theta=\frac{8}{23}$, (ef)~$\theta=\frac{17}{23}$. }
\label{PhysarumIntersectionProximity}
\end{center}
\end{figure}

Intersections of Physarum graphs for principle values of threshold $\theta$ with the Gabriel graph and the minimum spanning tree are shown in Fig.~\ref{PhysarumIntersectionProximity}.  

\begin{finding}
$\mathbf{GG}$/\{(Inuvik --- Wrigley), (Thunder Bay --- Radisson)\} $\subset$ $\mathbf{H}$  
\end{finding}

\begin{finding}
$\mathbf{MST} \subset \mathbf{P}(0)$
\end{finding}

'Raw' Physarum includes an 'ideal' acyclic spanning network $\mathbf{MST}$. This somehow characterises 
a good quality of a slime mould approximation of a transport network. The minimum spanning tree is not included 
in the high-threshold Physarum graph $\mathbf{P}(\frac{17}{23})$. However, there is a 'strong' component of $\mathbf{MST}$ which is included in the high-threshold Physarum graph (Fig.~\ref{PhysarumIntersectionProximity}). 
The strong component is a tree rooted in  the Toronto area. The tree's stem is  Toronto --- Winnipeg ---
 Saskatoon-Regina --- Edmonton. 
It has three branches. A small branch Winnipeg --- Thompson, and two end branches  Edmonton --- Calgary --- Vancouver area and Edmonton --- Yellowknife --- Wrigley --- Inuvik.

\section{Response to contamination}
\label{contamination}

\begin{figure}[!tbp]
\begin{center}
\subfigure[]{\includegraphics[width=0.39\textwidth]{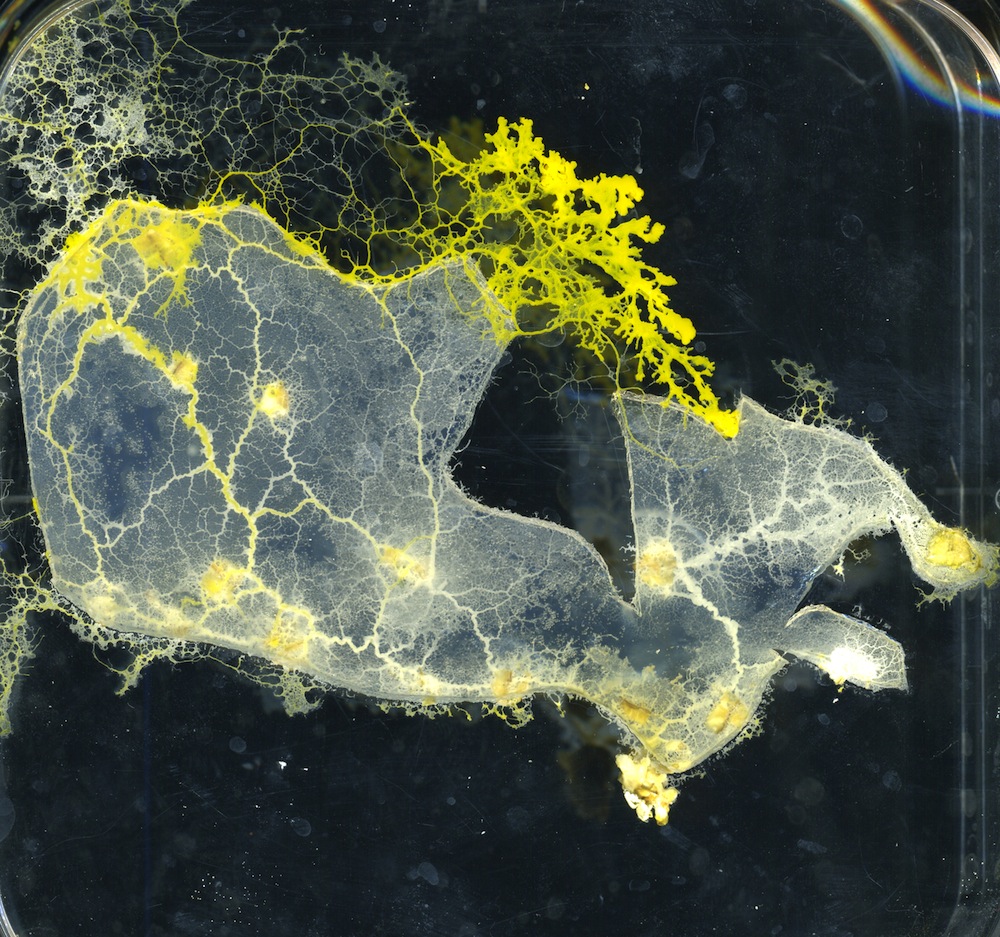}}
\subfigure[]{\includegraphics[width=0.39\textwidth]{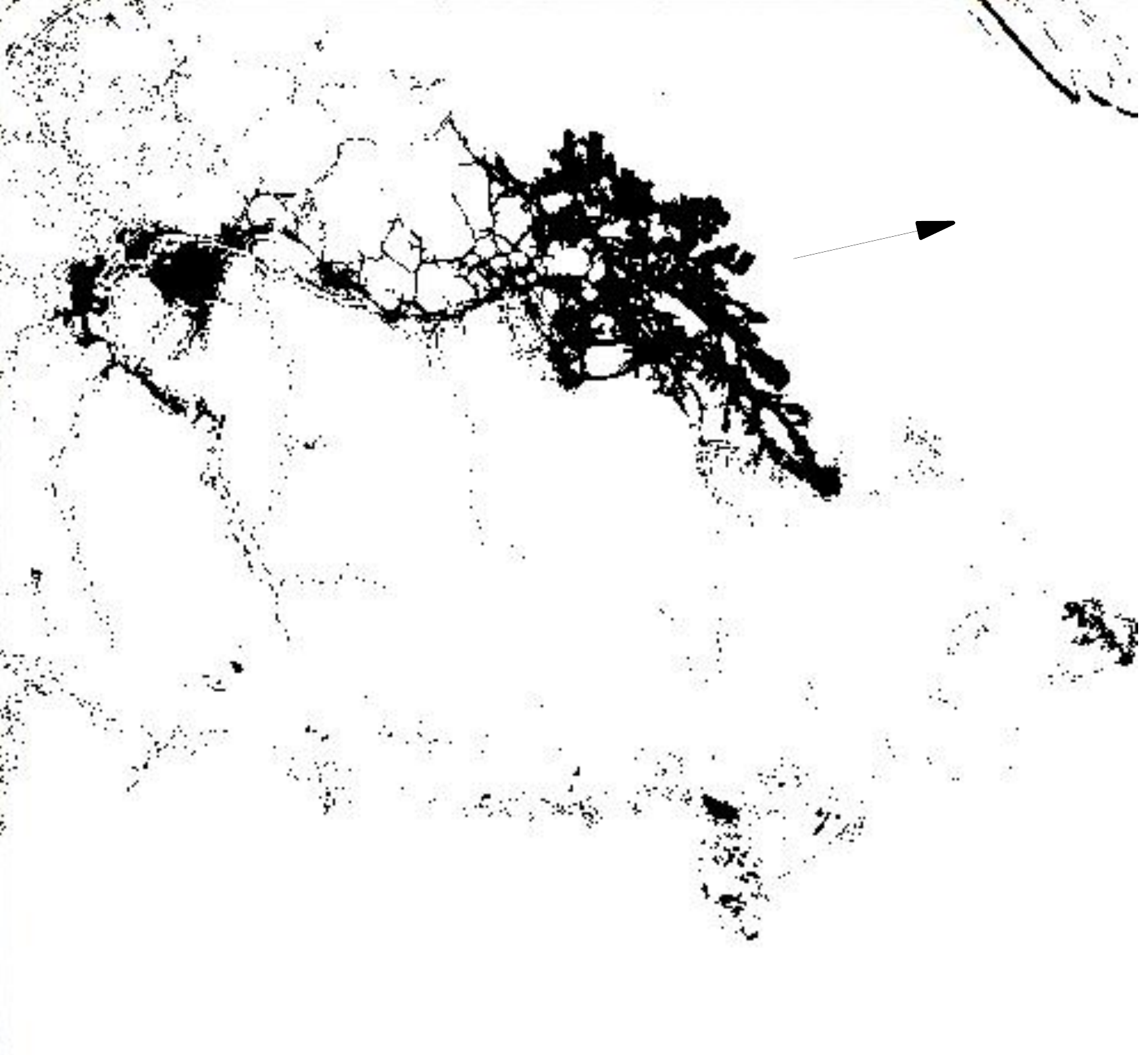}}
\subfigure[]{\includegraphics[width=0.39\textwidth]{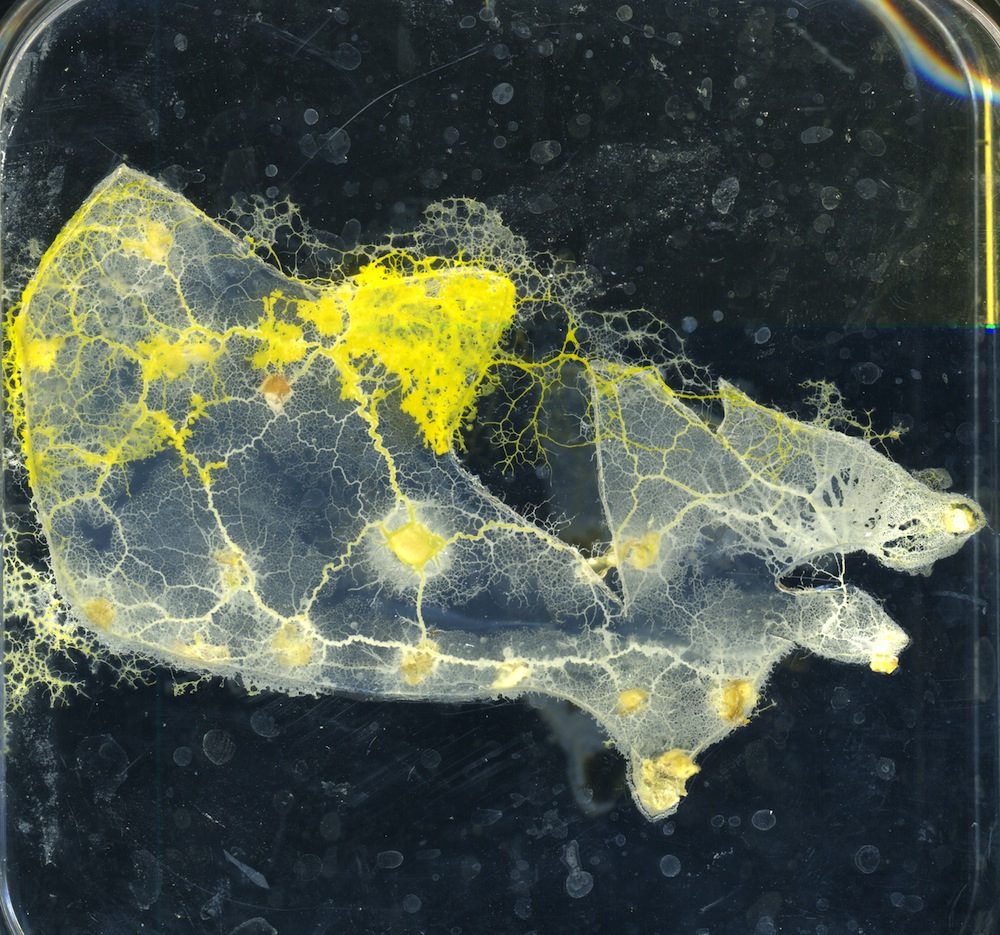}}
\subfigure[]{\includegraphics[width=0.39\textwidth]{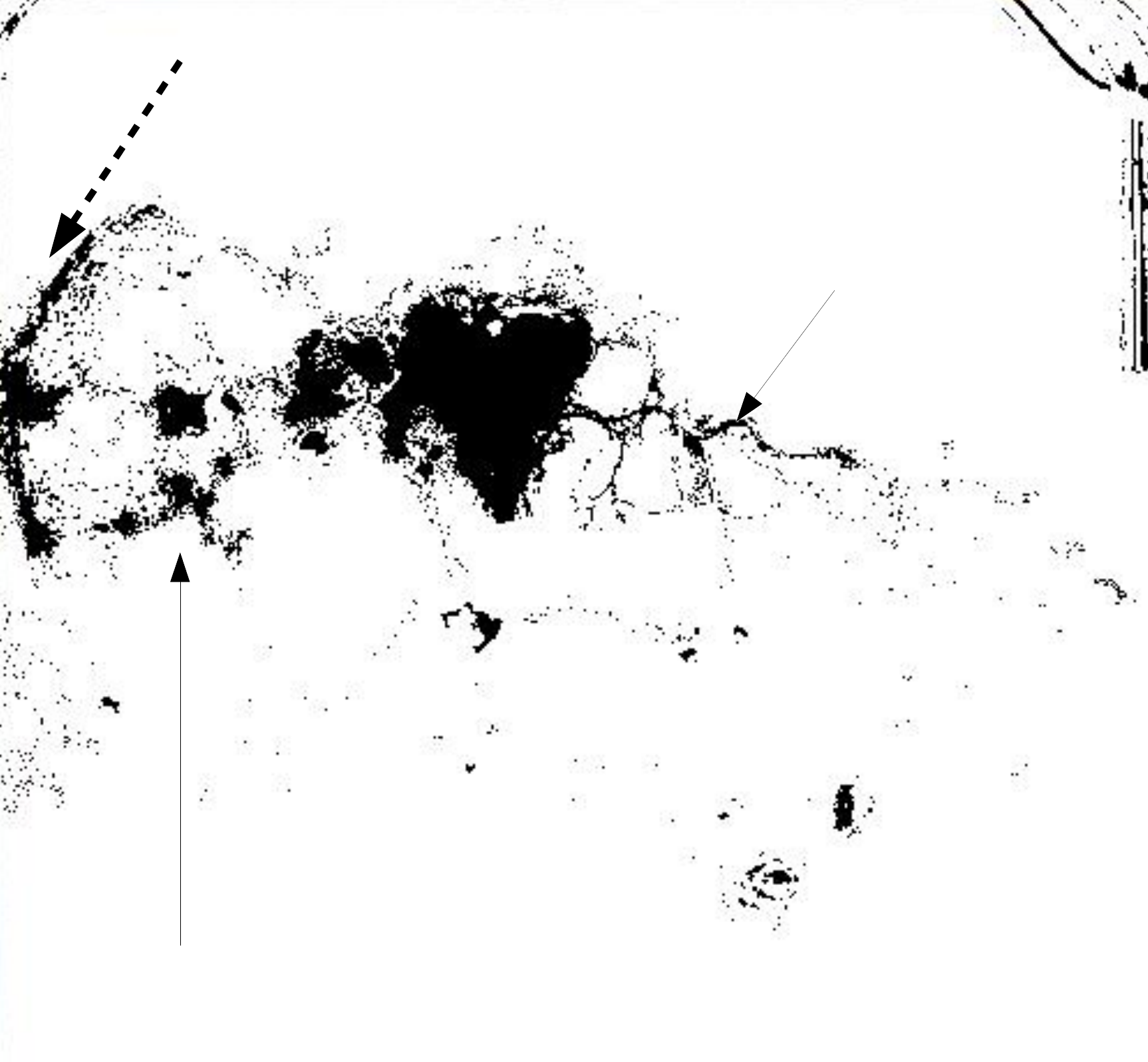}}
\subfigure[]{\includegraphics[width=0.39\textwidth]{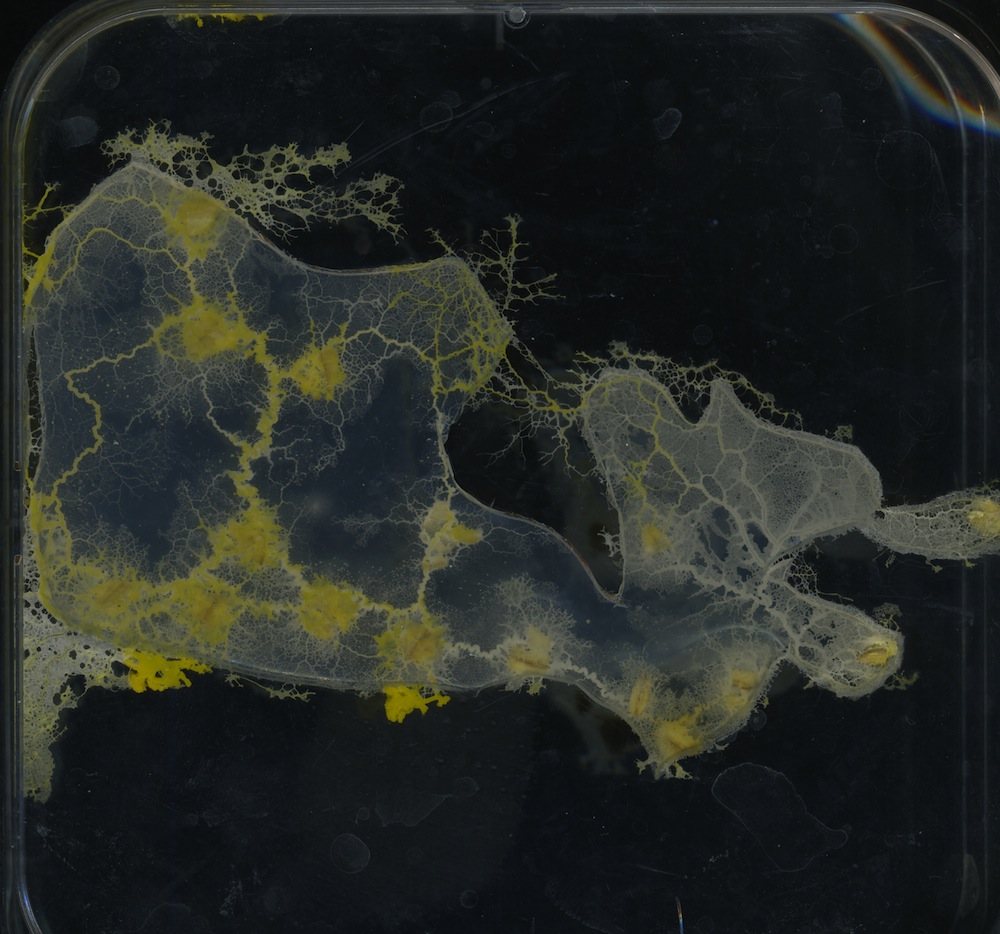}}
\subfigure[]{\includegraphics[width=0.39\textwidth]{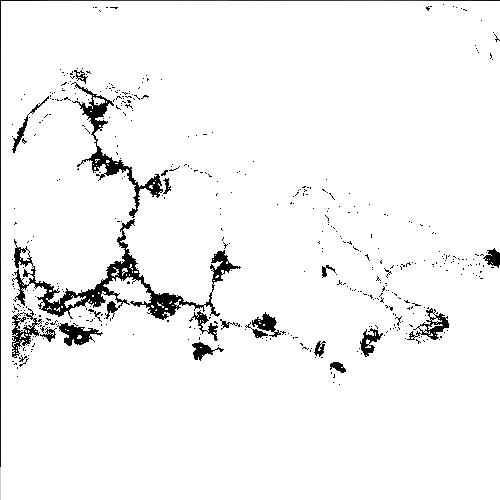}}
\caption{Physarum response to contamination. (ab)~Migration outside Canada.
(c--f)~Compensating activation of transport network unaffected by contamination.
(ace)~Original images. (bdf)~Binarized images: only pixels from (ace) which red  and green
components exceed 100 and blue component is less than 100 are drawn as black pixels in (bdf), 
otherwise white. 
}
\label{contamination1}
\end{center}
\end{figure}

\begin{figure}[!tbp]
\begin{center}
\subfigure[]{\includegraphics[width=0.39\textwidth]{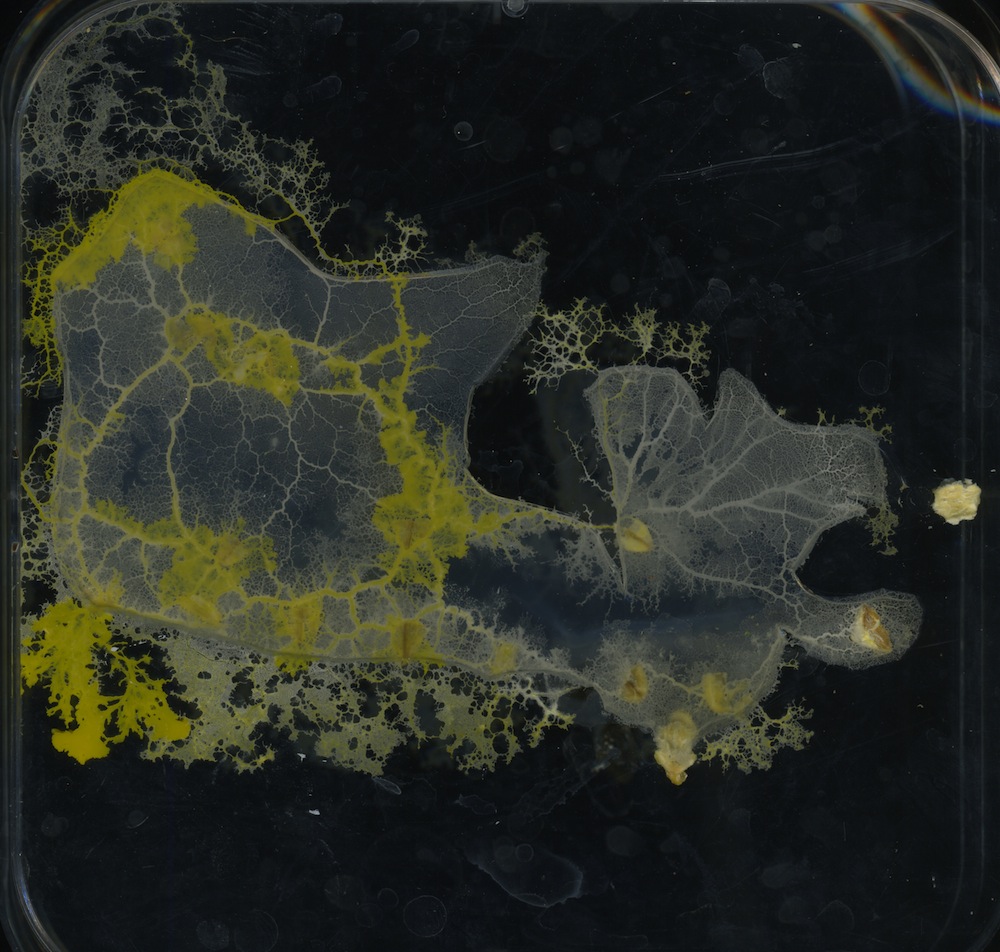}}
\subfigure[]{\includegraphics[width=0.39\textwidth]{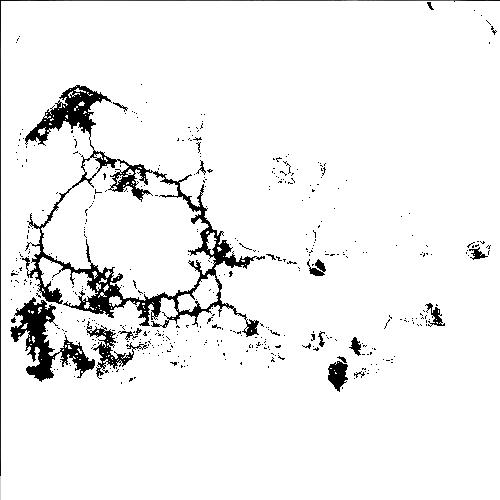}}
\subfigure[]{\includegraphics[width=0.39\textwidth]{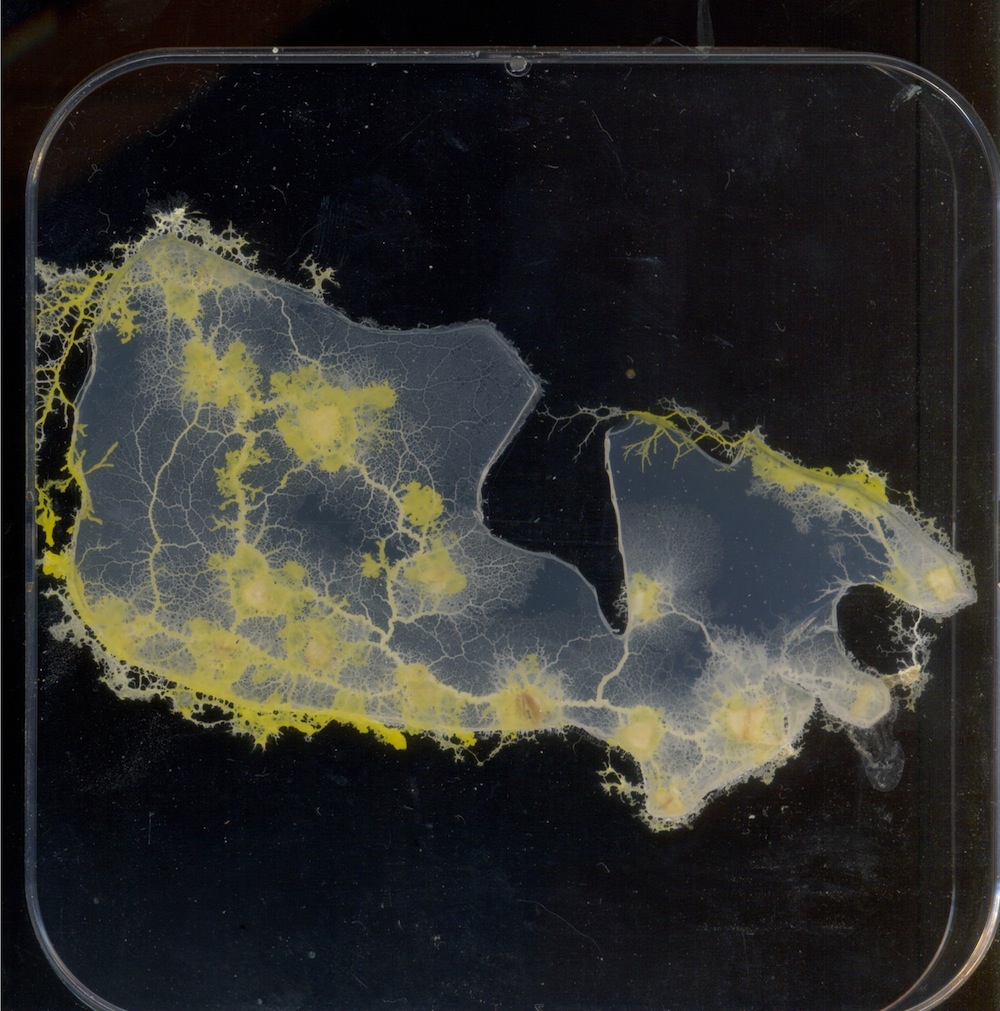}}
\subfigure[]{\includegraphics[width=0.39\textwidth]{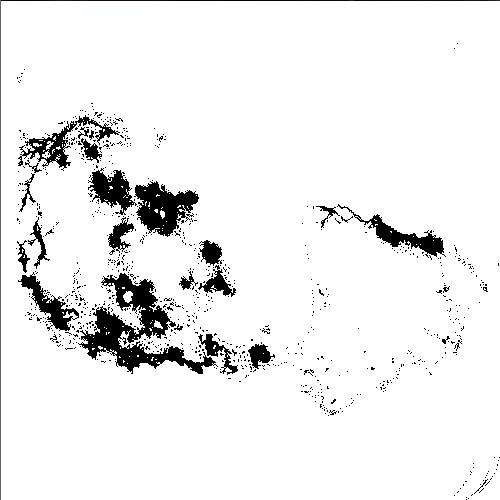}}
\caption{Physarum sprouting in response to contamination. 
(ab)~Sprouting from edges. (cd)~Sprouting from nodes. (ac)~Original colour images. (bd)~Binarized images.}
\label{contamination2}
\end{center}
\end{figure}

To imitate propagating contamination we place a crystal of sea salt (SAXA Coarse Sea Salt, a crystal weight around 
20~mg), in the place of Bruce Nuclear Power station. Inorganic salts are chemo-repellents for  
\emph{P. polycephalum} therefore sodium chloride diffusing in agar gel causes plasmodium to retreat from a contaminated zone.  We studied plasmodium's response circa 24~h after initiation of  contamination. During 24~h a contamination zone expands as far as Winnipeg an Thompson in the west and St. John's in the east. In some cases contamination spreads till Saskatoon-Moncton. In a few experiments plasmodium colony occupying St. Jonn's remains unaffected. 

\begin{finding}
In response to contamination propagating from Bruce Nuclear Power station, the plasmodium of \emph{P. polycephalum}
takes one or more of the following actions: migrates outside Canada, enhances the transport network outside the contaminated zone, sproutes indiscriminately from urban areas and transport links. 
\end{finding}  

The plasmodium's reactions are illustrated in Figs.~\ref{contamination1} and \ref{contamination2}. Four types of responses are observed in laboratory experiments.
\begin{itemize}
\item Plasmodium migrates outside Canada (Fig.~\ref{contamination1}ab). Typical waves of migration are from Nunavut towards Baffin Bay and Greenland and from British Columbia towards Washington and Oregon in USA. Due to the growth substrate being an agar plate cut in the shape of Canada, the plasmodium ends up on the bare plastic bottom of a Petri dish, therefore it does not migrate far away from Canada. 
\item Plasmodium enhances it foraging and colonisation activity in the parts unaffected by contamination  (Fig.~\ref{contamination1}c--f), mainly in Alberta, British Columbia, Northern Territories, Yukon and Nunavut. Protoplasmic tubes themselves are often increased in size and intensity of their colours, which reflects increased propagation of cytoplasm inside the tubes. For example, in Fig.~\ref{contamination1}cd plasmodium clearly shows hyper-activity in the Nunavut area, with the whole territory covered by spreading plasmodium.  Fig.~\ref{contamination1}cd illustrates hyper-activation of transport routes, particularly links Inuvik --- Wrigley, Yelloknife --- Wrigley, Edmonton --- Wrigley, Edmonton --- Vancouver area, Saskatoon-Regina --- Edmonton, and Winnipeg --- Saskatoon.
\item Plasmodium expands outside oat flakes and also produces processes protruding from the protoplasmic tubes
(Fig.~\ref{contamination2}a--d). This a common reaction of plasmodium in response to mechanical damage, e.g. cutting of protoplasmic tubes~\cite{PhysarumMachines}.  
\end{itemize}

\section{Discussion}

To imitate transport networks in Canada we represented major urban areas and transport nodes with oat flakes, 
inoculated plasmodium of \emph{Physarum polycephalum}, allowed the plasmodium to span all oat flakes with its network of protoplasmic tubes and analysed the structure of the protoplasmic network.  We found that in over 75\% of experiments \emph{P. polycephalum} an acyclic transport network consisting of a chain spanning urban areas along the south boundary of Canada, from Halifax-Moncton to Vancouver area, with branches Edmonton to Yellowknife to Wrigley to Inuvik and Yellowknife to Thompson.  In all experiments slime mould approximates all but Vancouver to Calgary links of Canadian highways networks. Both slime mould and Canadian highway networks have a strong spanning tree component and thus can be thought of as optimal transport networks. In laboratory experiments with slime mould we also detailed possible scenarios of transport network restructuring in a response to a spreading contamination. We believe our results make a substantial contribution towards nature-inspired analyses and design of human-made transport networks. Further experiments are necessary to determine how natural and geographical conditions, especially terrain, affect the exact topology of developing transport networks.

\end{document}